\documentclass[pre,preprint,showpacs,groupedaddress,repreint]{revtex4-1}

\usepackage{graphicx}% Include figure files
\usepackage{dcolumn}% Align table columns on decimal point
\usepackage{bm}% bold math
\usepackage{amsmath}
\usepackage{hyperref}% add hypertext capabilities
\usepackage{color}
\makeatother
\usepackage{tikz}
\begin{document}

%\preprint{}

\title{Transport of the moving barrier driven by chiral active particles}


%\homepage[]{}

%\thanks{}

\author{Jing-jing Liao$^{1,2}$}
\author{Xiao-qun Huang$^{1}$}
%\author{Wei-jing Zhu$^{1}$}
\author{Bao-quan  Ai$^{1}$}\email[Email: ]{aibq@scnu.edu.cn}
 %\email[Email: ]{wrzhong@jnu.edu.cn}
%\homepage[]{}

%\thanks{}
\affiliation{$^{1}$ Guangdong Provincial Key Laboratory of Quantum Engineering and Quantum Materials, School of Physics and Telecommunication
Engineering, South China Normal University, Guangzhou 510006, China.}
\affiliation{$^{2}$ College of Applied Science, Jiangxi University of Science and Technology, Ganzhou 341000, China.}

\date{\today}
\begin{abstract}
\indent Transport of a moving V-shaped barrier exposed to a bath of chiral active particles is investigated in a two-dimensional channel. Due to the chirality of active particles and the transversal asymmetry of the barrier position, active particles can power and steer the directed transport of the barrier in the longitudinal direction. The transport of the barrier is determined by the chirality of active particles. The moving barrier and active particles move in the opposite directions. The average velocity of the barrier is much larger than that of active particles. There exist optimal parameters (the chirality, the self-propulsion speed, the packing fraction and the channel width) at which the average velocity of the barrier takes its maximal value. In particular, tailoring the geometry of the barrier and the active concentration provides novel strategies to control the transport properties of micro-objects or cargoes in an active medium.
\end{abstract}

%\pacs{05. 40. Fb, 02. 50. Ey, 05. 40. -a}
%\keywords{ chiral active particles, transport, V-shape barrier }



\maketitle
\section {Introduction}
\indent In recent years, active particle transport in complex environments has
attracted widely attention and much interest in biology, chemistry and
nanotechnology \cite{Marchetti,Elgeti,ref3}. Different from the features of passive
colloids, the intrinsic nonequilibrium property of active particles, also
known as self-propelled Brownian particles or microswimmers and
nanoswimmers, can take energy from their environment and produce a force
which pushes them forward \cite{Ramaswamy}. When
an active particle features a common symmetry axis of body and
self-propelling force, it swims linearly only \cite{Howse}. Otherwise it experiences a constant torque, which is called
'chiral active particle', and performs circular motion in two dimensions and
helicoidal motion in three dimensions due to the self-propulsion force being
not aligning with the propulsion direction \cite{Van}. Chiral active matter can exhibit intriguing
phenomena \cite{Golestanian,Liebchen,Vicsek,Levis,Potiguar,Galajda,Kaiser2,Kaiser3,Kaiser,Di,Wan,Ghosh,Angelani,Pototsky,Ghosh2,Koumakis,Rusconi,Buttinoni,Schwarz,Stenhammar,Fily,Stenhammar2,Nepusz,Ai1,Ai2,Ten,Engel,Enculescu,Guidobaldi}, such as self-organization and collective behaviors. Instances of such new
active matter system can be found in active chiral fluids \cite{Tjhung,MLeoni,Furthauer,ELushi} and many biological micro-swimmers ranging
from spermatozoa \cite{Friedrich} and Escherchia Coli \cite{Diluzio,Di2} to
Listeria monocytogenes \cite{Shenoy}.

\indent In particular, the behaviors and dynamics of obstacles exposed to an active
fluid and the transport properties of active particles take on great
importance in several practical applications \cite{Di,Galajda,Kaiser,Wan,Angelani,Ghosh,Kaiser2,Kaiser3,Angelani2}, such
as driving microscopic gears and motors \cite{Di,Angelani}, the capture and
rectification of active particles \cite{Galajda,Wan,Ghosh,Kaiser2,Kaiser3}, and using active
suspensions to propel wedge-like carriers \cite{Kaiser,Angelani2}. The interactions of
active particles with obstacles have been investigated by using theoretical
studies, simulations and experiments \cite{Galajda,Potiguar,Kaiser2,Kaiser3,Kaiser,ref45,ref46,ref47,ref48,Angelani2,Mallory,Smallenburg,Marini,Hu,Wu}. Potiguar and
coworkers \cite{Potiguar} found a vortex-type motion of self-propelled particles around convex symmetric obstacles and an steady
particle current in an array of non-symmetric convex obstacles. Galajda et al. \cite{Galajda} observed a ratchet motion of the swimming bacteria placed in a confined area containing an array of funnel shapes. Kaiser et al. \cite{Kaiser2} showed the interaction between active self-propelled rods
and stationary wedges was as a function of the wedge angle. In a subsequent work, Kaiser et al. \cite{Kaiser} demonstrated that the directed transport of mesoscopic
carriers through the suspension could be powered and steered by collective turbulentlike motion in a bacterial bath. C. Reichhardt and C. J. O. Reichhardt \cite{ref45} found that a ratchet effect produced by chirality was observed for circularly moving particles interacting with a periodic array of asymmetric L-shaped obstacles. They also studied active particles which are placed in an asymmetric array of funnels could produce a ratchet effect even in the absence of an external drive \cite{ref46}. Ratchet reversals produced by collective effects and
the use of active ratchets to transport passive particles were investigated and reviewed in Ref. \cite{ref47} and Ref. \cite{ref48}. Angelani et al. \cite{Angelani2} showed active particles powered the asymmetric arrow-shaped barriers to move in one dimension. Mallory and coworkers \cite{Mallory} numerically studied the transport of a asymmetric tracer immersed in a high dilution suspension of self-propelled nanoparticles.  Marconi et al. \cite{Marini} studied the role of self-propulsion in active particles interacting with
a moving semipermeable membrane with a constant velocity. Chiral active particles can be rectified in the longitudinal direction when the potentials or the fixed obstacles are asymmetric along the transversal direction in the periodic channel \cite{Hu,Wu}.

\indent In the previous studies, active particles are considered to interact with a fixed obstacle, or an obstacle with a constant velocity or a non-symmetric obstacle driven by active particles. However, the directed transport of the moving symmetric barrier driven by chiral active particles has not been considered yet, which results in a fascinating wealth of new nonequilibrium phenomena actually. In this paper, we expose a V-shaped barrier to a bath of chiral active particles. We emphasize on studying the interplay between active particles and the barrier, finding the directed transport of the barrier powered and steered by active particles and investigating how the system parameters and the moving barrier affect the rectification of chiral active particles. We also focus on comparing the transport of chiral active particles between the cases of the barrier fixed and moving. It is found that the transport speed of the barrier is much larger than that of active particles. The scaled average velocity of active particles in the moving case is reduced much than in the fixed case. The moving barrier and active particles move in the opposite directions. We can obtain maximal scaled average velocity of the barrier when the system parameters are optimized. Our results can be applied practically in powering and steering carriers and motors by a bath of bacteria or artificial microswimmers.

\section{Model and Methods}

\indent We consider $n_a$ chiral active particles with radius $r$ moving in a two-dimensional straight channel with periodic boundary conditions (the period $L_{x}$ ) in the $x$-direction and hard wall boundary conditions (the width $L_{y}$) in the $y$-direction as shown in Fig. 1. A V-shaped barrier with angle $\alpha$ is exposed to the bottom of the channel. In order to restrict the V-shaped barrier moving only along the $x$-direction, two parallel tracks (active particles cannot feel the tracks) are settled in the channel, one is fixed at the bottom of channel and the other is fixed at the top of the barrier. Each side of the V-shaped barrier consists of $n_{p}$ particles with radius $r$. The total particle number of chiral active particles and the V-shaped barrier is $N=n_a+2n_p+1$. The position of particle $i$ is described by $\bf{r_i}\equiv$$(x_i,y_i)$, and its speed direction is denoted by the orientation $\theta_i$ of the polar axis $\mathbf{n_i}\equiv(\cos\theta_i,\sin\theta_i)$. We define $\mathbf{F}_i=F^{x}_{i}\mathbf{e}_x+F^{y}_{i}\mathbf{e}_y=\sum_{j}\mathbf{F}_{ij}$ and $\mathbf{G}_{i}=G^{x}_{i}\mathbf{e}_x+G^{y}_{i}\mathbf{e}_y=\sum_{j}\mathbf{G}_{ij}$ as the forces acting on particle $i$ from other active particles and from the V-shaped barrier, respectively. The particle $i$ obeys the following overdamped Langevin equations:

\begin{figure}[htbp]
\begin{center}
\includegraphics[width=12cm]{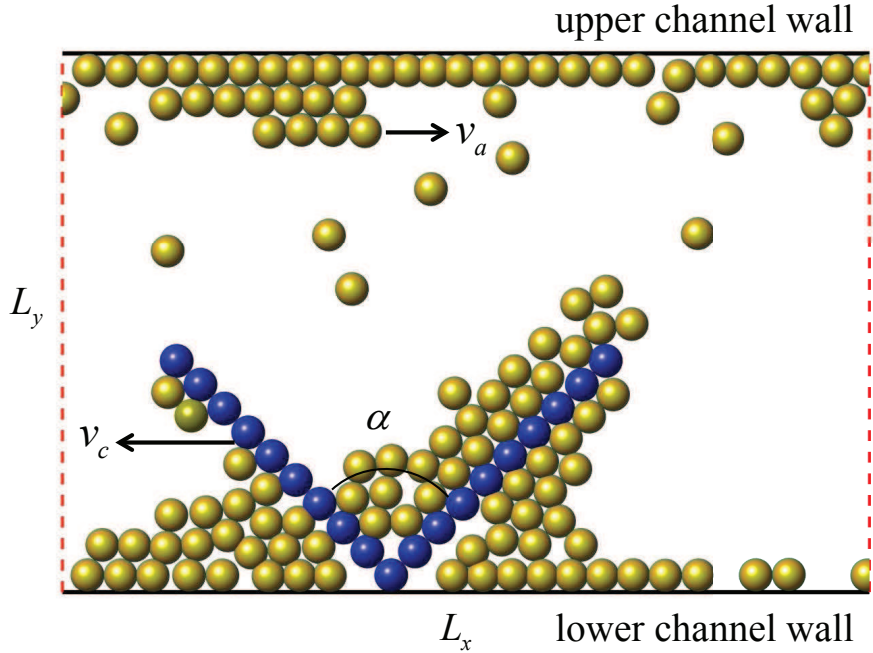}
\caption{Schematic of chirality-powered motor. A V-shaped barrier with
angle $\alpha$ is exposed to the bottom of the
channel. The barrier consists of ($2n_p+1$) particles with
radius $r$. The V-shaped barrier has two cases: fixed and moving. Periodic boundary conditions are imposed in
the $x$-direction, and hard wall boundaries in the $y$-direction. $v_a$ and $v_c$ denote the average velocity of chiral active particles and the center of the V-shaped barrier along the $x$-direction, respectively.}
\end{center}
\end{figure}

\begin{equation}\label{e1}
\frac{dx_i}{dt}=\mu[F^x_i+G^x_i]+v_0\cos\theta_i+\sqrt{2D_0}\xi_{i}^{x}(t),
\end{equation}
\begin{equation}\label{e2}
\frac{dy_i}{dt}=\mu[F^y_i+G^y_i]+v_0\sin\theta_i+\sqrt{2D_0}\xi_{i}^{y}(t),
\end{equation}
\begin{equation}\label{e3}
\frac{d\theta_i}{dt}=\Omega+\sqrt{2D_{\theta}}\xi_{i}^{\theta}(t),
\end{equation}
where $v_{0}$ denotes the magnitude of self-propelled velocity and $\mu$ is the mobility. $\Omega$ is the angular velocity and its sign determines the chirality of active particles. Particles are defined as the clockwise particles for $\Omega<0$ and the counterclockwise particles for $\Omega>0$. The translational and rotational diffusion coefficients are denoted by $D_{0}$ and $D_{\theta}$, respectively. $\xi_{i}^{x}(t)$, $\xi_{i}^{y}(t)$, and $\xi_{i}^{\theta}(t)$ are the unit-variance Gaussian white noises with zero mean.

\indent  The force $\mathbf{F}_{ij}$ between active particle $i$ and $j$, and the force $\mathbf{G}_{ij}$ between active particle $i$ and the barrier particle $j$ are taken as the linear spring form with the stiffness constant $k_1$ and $k_2$, respectively. $\mathbf{F}_{ij}=k_1(2r-r_{ij})\mathbf{e}_{r}$, if $r_{ij}<2r$ ($\mathbf{F}_{ij}=0$ otherwise), and $r_{ij}$ is the distance between active particle $i$ and $j$. $\mathbf{G}_{ij}=k_2(2r-r_{ij})\mathbf{e}_{r}$, if $r_{ij}<2r$ ($\mathbf{G}_{ij}=0$ otherwise), and $r_{ij}$ is the distance between active particle $i$ and the barrier particle $j$. We use large values of $k_1$ and $k_2$ to imitate hard particles. It ensures that particle overlaps decay quickly.

\indent We can rewrite Eqs.(1)-(3) in the dimensionless forms by introducing the characteristic length scale and the time scale: $\hat{x}=\frac{x}{r}$, $\hat{y}=\frac{y}{r}$, and $\hat{t}=\mu kt$,

\begin{equation}\label{e4}
\frac{d\hat x_i}{d\hat t}=\hat F_{i}^x+\hat G_{i}^x+\hat v_0\cos\theta_i+\sqrt{2\hat D_0}\hat \xi_{i}^{x}(\hat t),
\end{equation}
\begin{equation}\label{e5}
\frac{d\hat y_i}{d\hat t}=\hat F_{i}^y+\hat G_{i}^y+\hat v_0\sin\theta_i+\sqrt{2\hat D_0}\hat \xi_{i}^{y}(\hat t),
\end{equation}
\begin{equation}\label{e6}
\frac{d\theta_i}{d\hat t}=\hat\Omega+\sqrt{2\hat D_{\theta}}\hat \xi_{i}^{\theta}(\hat t),
\end{equation}
and the other parameters can be rewritten as $\hat{L}_x=\frac{L_x}{r}$, $\hat{L}_y=\frac{L_y}{r}$,$\hat{v}_0=\frac{v_0}{\mu kr}$, $\hat{D}_0=\frac{D_0}{\mu kr^2}$, and $\hat{D}_\theta=\frac{D_\theta}{\mu k}$. In the following discussions, only the dimensionless variables will be used, and the hats for all quantities appearing in the above equations shall be omitted .

\indent By integration of the Langevin Eqs.(4)-(6) using the second-order stochastic Runge-Kutta algorithm, we can get the transport behaviors of the quantities. To quantify the ratchet effect, we only calculate average velocity in the $x$-direction because particles along the $y$-direction are confined and directed transport only occurs in the $x$-direction. In the asymptotic long-time regime, we can obtain the average velocity of chiral active particles along the $x$-direction using the following formula
\begin{equation}\label{e7}
v_a=\frac{1}{n_a}\sum_{i=1}^{n_a}\lim_{t\rightarrow\infty}\frac{x_{i}(t)-x_{i}(0)}{t}.
\end{equation}

\indent The forces acting on the V-shaped barrier particle $j$ from the chiral active particle $i$ are defined as $\mathbf{G}_{j}=G^{x}_{j}\mathbf{e}_x+G^{y}_{j}\mathbf{e}_y=\sum_{i}\mathbf{G}_{ij}$. It leads to the barrier moving when the V-shaped barrier is not fixed. The motion equation for the center of the V-shaped barrier is as follows:
\begin{equation}\label{e8}
\frac{dx_c}{dt}=\gamma G^x,
\end{equation}
where $\gamma$ is the coefficient we set, the barrier is fixed when $\gamma=0$ and can be driven to move along the $x$-direction when $\gamma=1.0$. $x_c$ is the center position of the barrier in the $x$-direction , $G^x=\sum_{j}G^{x}_{j}/(2n_p+1)$ is the average force acting on the center of the V-shaped barrier along the $x$-direction. In the asymptotic long-time regime, the average velocity of the center of the V-shaped barrier along the $x$-direction can be obtained from the following formula
\begin{equation}\label{e9}
v_c=\lim_{t\rightarrow\infty}\frac{x_{c}(t)-x_{c}(0)}{t}.
\end{equation}

\indent We define the ratio between the area occupied by particles and the total available area as the packing fraction $\phi=\pi(2n_p+1+n_a)r^2/(L_x L_y)$.  In addition, we define the scaled average velocity as $\eta_a=v_a/v_0$ and $\eta_o=v_c/v_0$ which respectively stand for rectification of chiral particles and the barrier.

\section {Numerical results and discussion}
\indent In our simulations, we have considered more than 100 realizations to improve accuracy and minimize statistical errors. The total integration time was chosen to be more than $10^6$ and the integration step time was smaller than $10^{-3}$. Unless otherwise noted, our simulations are under the parameter sets: $L_x=24.0$ and $L_y=16.0$. We vary $\Omega$, $D_0$, $D_\theta$, $n_p$, $v_0$, $\alpha$, $\phi$, and $L_y$ to calculate average velocity of chiral active particles and the V-shaped barrier when the barrier is fixed and moving.

\indent Actually, there are two critical elements of ratchet setup in nonlinear systems \cite{Denisov}. One is (a) Asymmetry (temporal and/or spatial), which can violate the left-right symmetry of the response. The other is (b)
fluctuating input zero-mean force: it should break thermodynamical equilibrium, which forbids a directed transport appearing due to the second law of thermodynamics. For our system, the asymmetry comes from the
upper-lower asymmetry of the channel due to the position of the V-shaped barrier and the fluctuating input zero-mean force comes from the self-propulsion of active particles. Because the circular trajectory radius of chiral particles $v_0/\left| \Omega \right|$ is much larger than the channel cell, chiral particles slide along the walls rather than move circularly. The channel is upper-lower asymmetric, therefore, the motion time along the upper wall is
significantly smaller than along the lower wall. The counterclockwise particles $\Omega >0$ on average move to the left, and the clockwise particles $\Omega <0$ on average move to the right.

\begin{figure}[htpb]
\vspace{1cm}
  \label{fig2}\includegraphics[width=0.45\columnwidth]{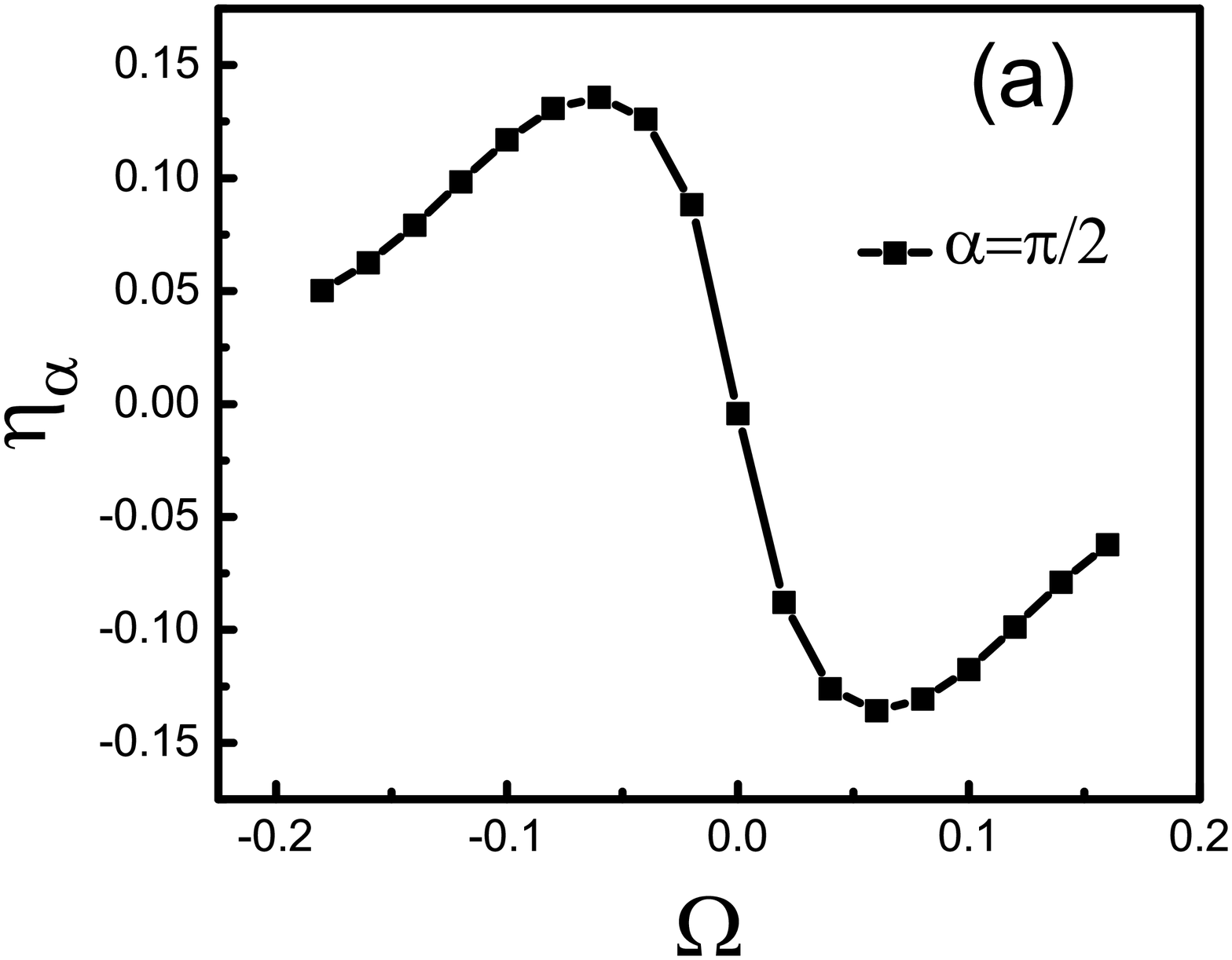}
  \includegraphics[width=0.45\columnwidth]{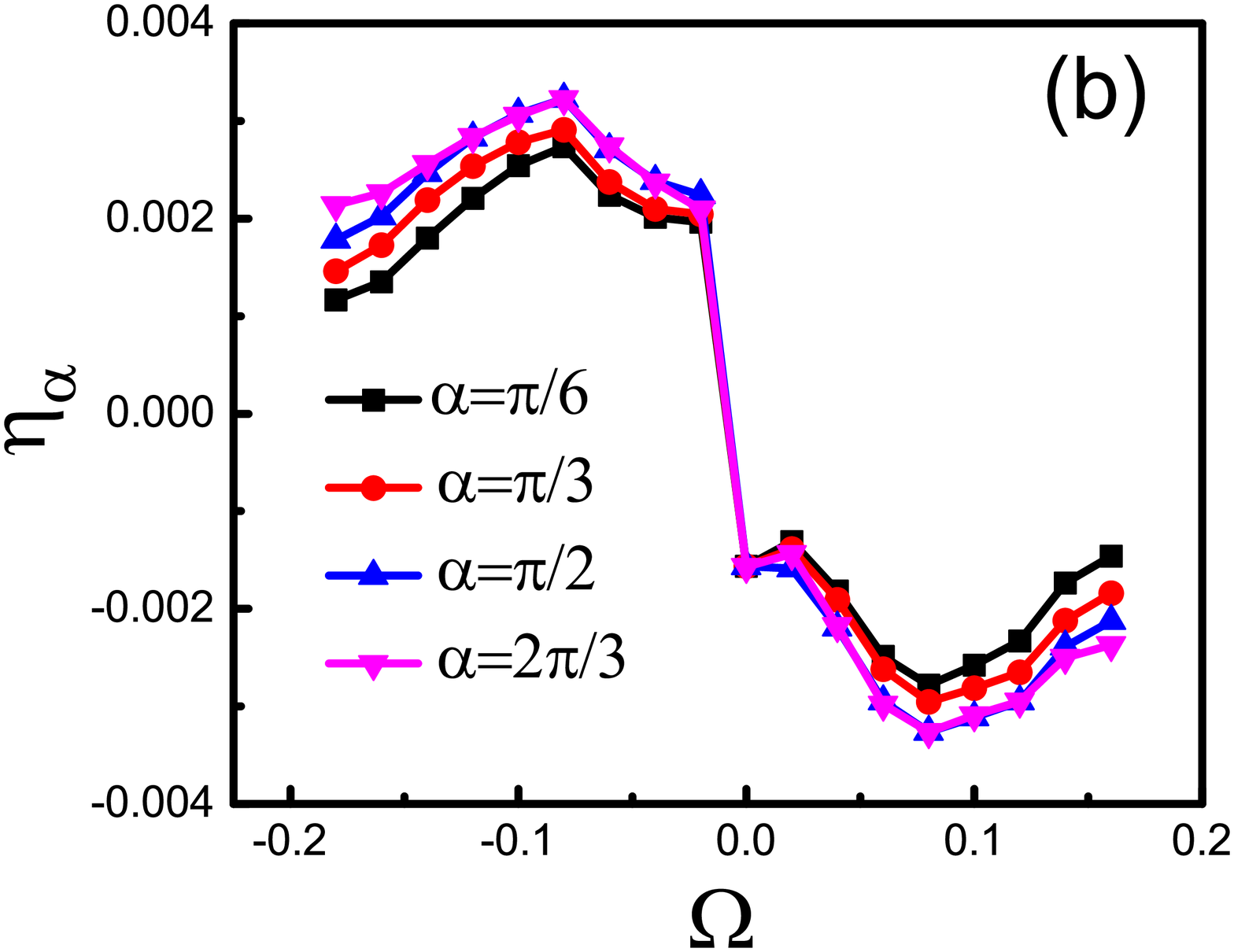}
  \includegraphics[width=0.45\columnwidth]{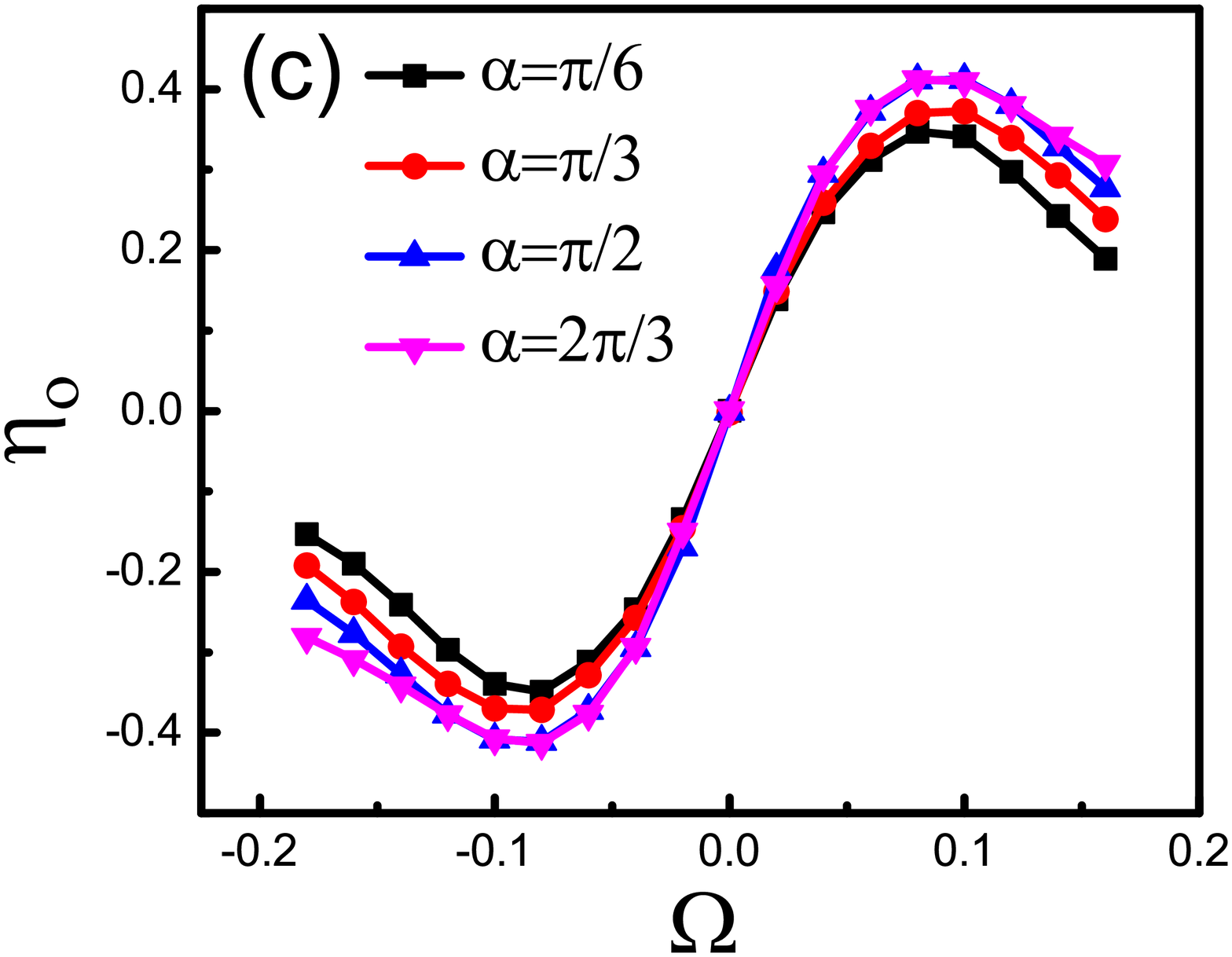}
  \caption{The scaled average velocity $\eta_a$ and $\eta_o$ as a function of the angular velocity $\Omega$. (a) Active particles for the fixed case at $\alpha=\pi/2$ .
  (b) Active particles for the moving case at $\alpha=\pi/6$, $\pi/3$, $\pi/2$, and $2\pi/3$. (c) The moving barrier at $\alpha=\pi/6$, $\pi/3$, $\pi/2$, and $2\pi/3$.
  The other parameters are $D_0=0.1$, $D_\theta=0.01$, $v_0=1.0$, $N=150$, $L_y=16.0$, and $n_p=9$. }
\end{figure}

\indent Figure 2 shows the scaled average velocity as a function of the angular
velocity $\Omega $. When the V-shaped barrier is fixed, the average velocity
of the obstacle is zero. For active particles (see Fig. 2(a)), $\eta_a$ is
negative for $\Omega >0$, zero at $\Omega =0$, and positive for $\Omega <0$.
The sign of $\Omega$ completely determine the transport direction of active
particles. That is to say, we can separate active particles with different
chiralities due to their different directions of motion. Additionally, when
$\Omega \to 0$, the chirality can be neglected, and the ratchet effect
disappears because the symmetry of the system cannot be broken, thus $\eta
_a \to 0$. When $\Omega \to \infty $, the self-propelled angle changes very
fast, particles will experience a zero averaged force, so $\eta_a $ tends to
zero. Therefore, there exists an optimal value of $\left| \Omega \right|$ at
which $\eta_a $ takes its maximal value.

\indent When the V-shaped barrier can move, the scaled average velocity of active
particles (see Fig. 2(b)) is reduced much than that in the fixed case, while the
transport speed of the barrier (see Fig. 2(c)) is about much larger than that of
active particles. The movement direction of the obstacle is also completely
determined by the sign of $\Omega $ and is opposite to the direction of active particles. The transport
behaviors which are the same as the above are demonstrated in the following
results (see Fig. 2-Fig. 8). Now we explain the underlying reason for the
barrier and chiral particles transport. The nonequilibrium driving which
comes from the chiral particles breaks thermodynamical equilibrium and power
the V-shaped barrier to move in the $x$-direction. Because the driving forces on
the barrier come from chiral particles, their transport behavior are similar
and in opposite directions. In our simulation, we choose $n_p =9$ and
$N=150$. In other words, the barrier consists of 19 particles and there are
131 chiral active particles. Similar to the collision between a large mass
of moving object and a small mass of stationary object, all self-propelled particles ($n_a=131$)
act on the barrier ($n_p=9$) resulting in much larger transport speed of
the barrier and much smaller velocity of active particles. Additionally, the velocity of the barrier is about 131 times larger than that of active particles. That is to say, the velocity ratio between the barrier and active particles is decided by the number of active particles. The force acting on the center of the barrier increases as the increasing number of active particles, then the velocity ratio between the barrier and active particles increases. When $\Omega \to 0$ and $\Omega \to \infty$, the scaled average velocity of active particles $\eta_a \to 0$, thus the driving effect can be neglected and the scaled average velocity of
the barrier $\eta _o$ goes to zero. Therefore, there exists an optimal
value of $|\Omega|$ at which $\eta_o$ takes its maximal
value. Additionally, we can control the movement direction of the barrier by tuning the angular
velocity of chiral particles which is a new technique and advantage in contrast to using achiral particles.

\begin{figure}[htpb]
\vspace{1cm}
  \label{fig3}\includegraphics[width=0.45\columnwidth]{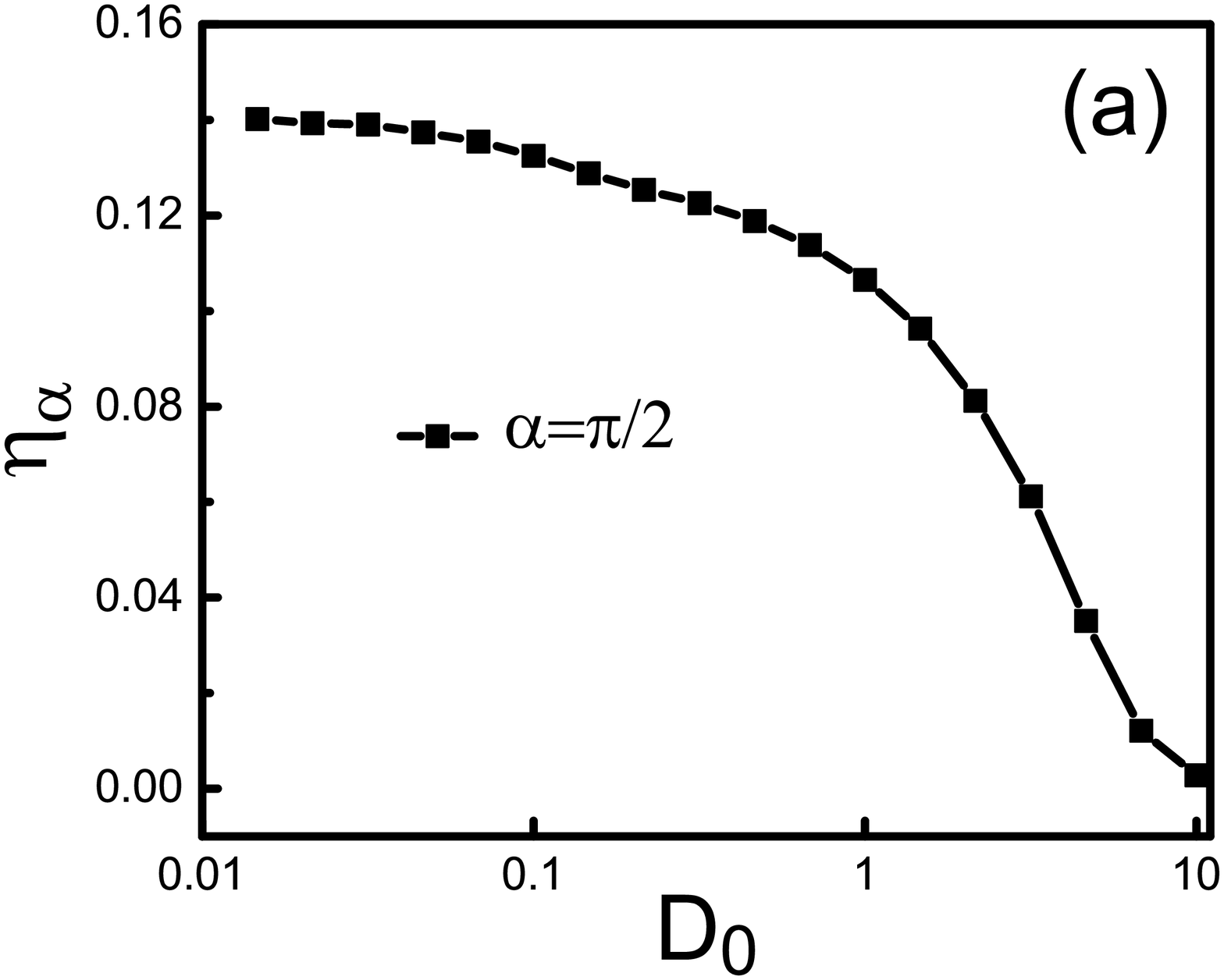}
  \includegraphics[width=0.45\columnwidth]{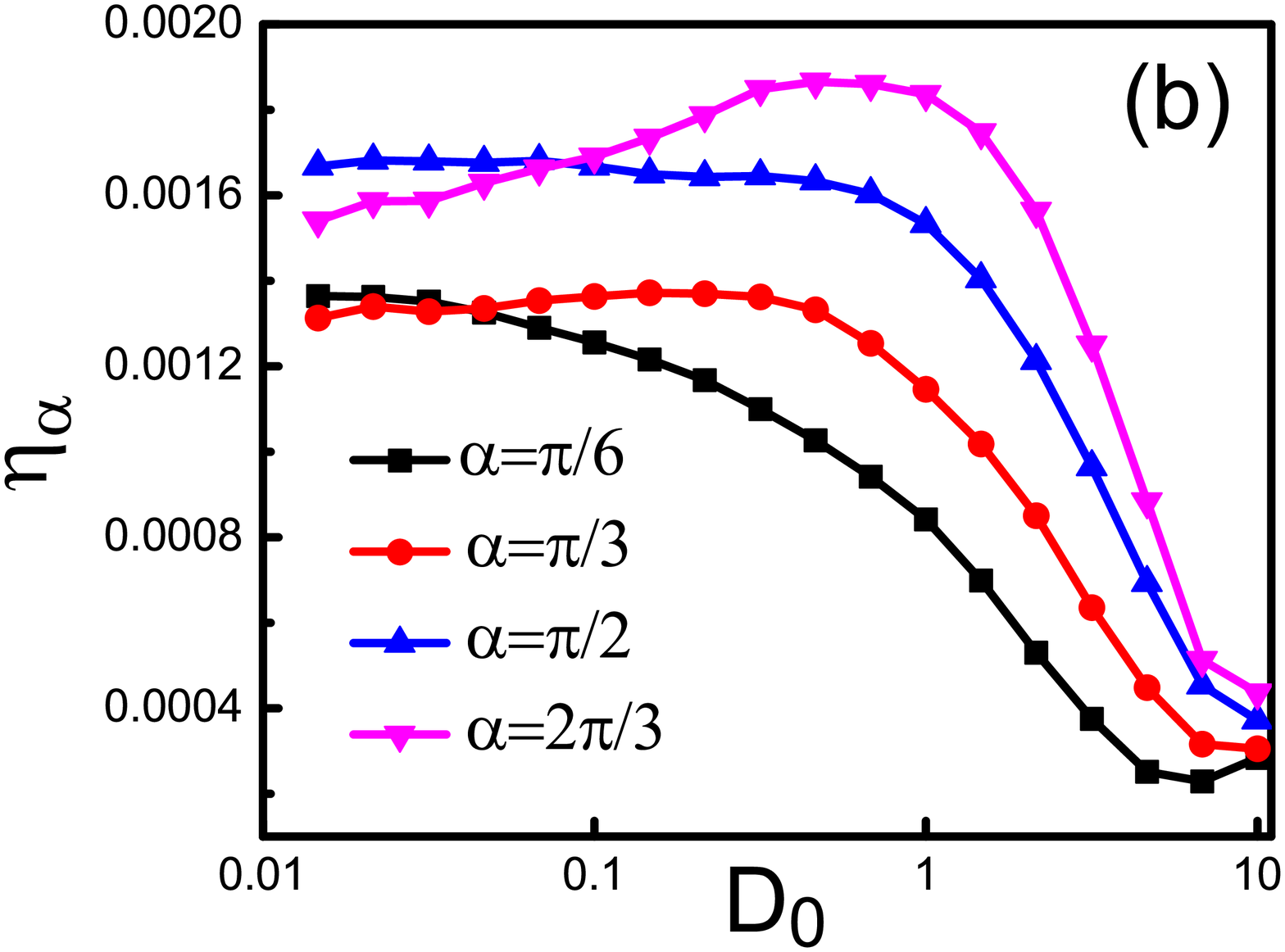}
  \includegraphics[width=0.45\columnwidth]{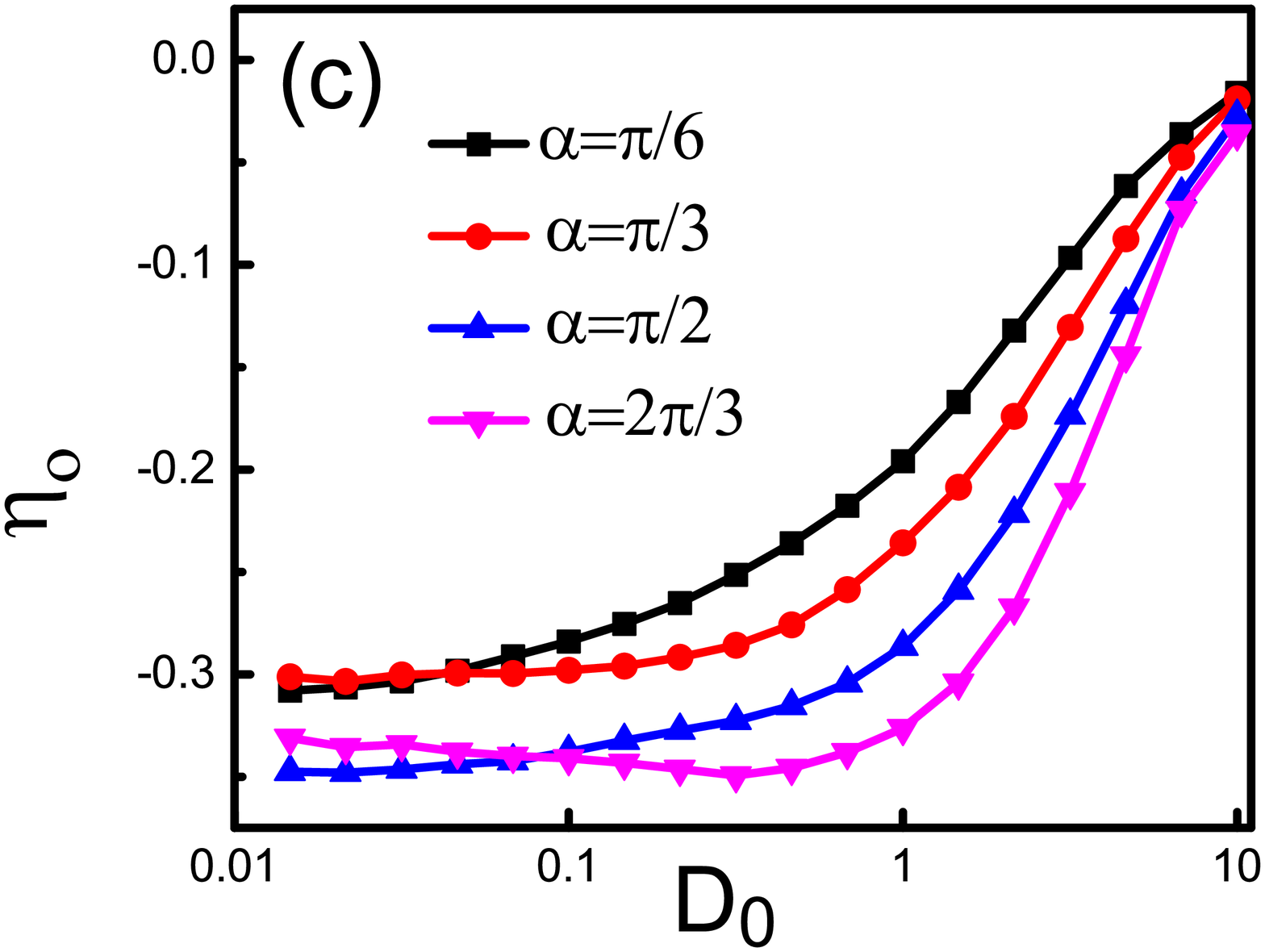}
  \caption{The scaled average velocity $\eta_a$ and $\eta_o$ as a function of the translational diffusion coefficient $D_0$. (a) Active particles for the fixed case at $\alpha=\pi/2$.
  (b) Active particles for the moving case at $\alpha=\pi/6$, $\pi/3$, $\pi/2$, and $2\pi/3$. (c) The moving barrier at $\alpha=\pi/6$, $\pi/3$, $\pi/2$, and $2\pi/3$.
  The other parameters are $\Omega=-0.05$, $D_\theta=0.01$, $v_0=1.0$, $N=150$, $L_y=16.0$, and $n_p=9$. }
\end{figure}

\indent Figure 3 displays the scaled average velocity versus the translational
diffusion coefficient $D_0$. As we know, the translational diffusion coefficient $D_0$ can cause
two results: (A) reducing the self-propelled driving which blocks the
ratchet transport when the particles can easily pass across the barrier. (B)
Facilitating particles to cross the barrier which promotes the
rectification when the particles cannot easily stride over the barrier. When
the barrier is fixed (see Fig. 3(a)), active particles can easily cross the barrier, the factor A dominates the transport, thus the rectification $\eta_a$ decreases monotonically with increasing $D_0$. Compared with the fixed case, active particles cannot easily stride over the barrier when the barrier can move.
When $D_0$ increases from zero, the factor A firstly dominates the transport at $\alpha=\pi/6$, $\pi/3$, and $\pi/2$ and the average velocity decreases, while the factor B firstly dominates the transport at $\alpha=2\pi/3$  and the average velocity increases.
This is because the larger the angle $\alpha$ is, the more particles are trapped in the corner of the barrier. When $D_0 \to\infty$, the translational diffusion is very large, the effect of the asymmetric barrier disappears and the scaled average velocity $\eta_a$ goes to zero. Therefore, in the moving case, the scaled average velocity $\eta_a$ decreases
monotonously with increasing $D_0$ at $\alpha=\pi/6$, $\pi/3$, and $\pi/2$, while there exists
an optimal $D_0$ value where the rectification is maximal at $\alpha=2\pi/3$ (see Fig. 3(b)). Similarly, on increasing $D_0$ from zero, the magnitude $|\eta_o|$ of the average velocity of the moving barrier decreases
monotonously at $\alpha=\pi/6$, $\pi/3$, and $\pi/2$, while  the magnitude $|\eta_o|$ is a peaked function of $D_0$ at $\alpha=2\pi/3$ (shown in Fig. 3(c)).

\begin{figure}[htpb]
\vspace{1cm}
  \label{fig4}\includegraphics[width=0.45\columnwidth]{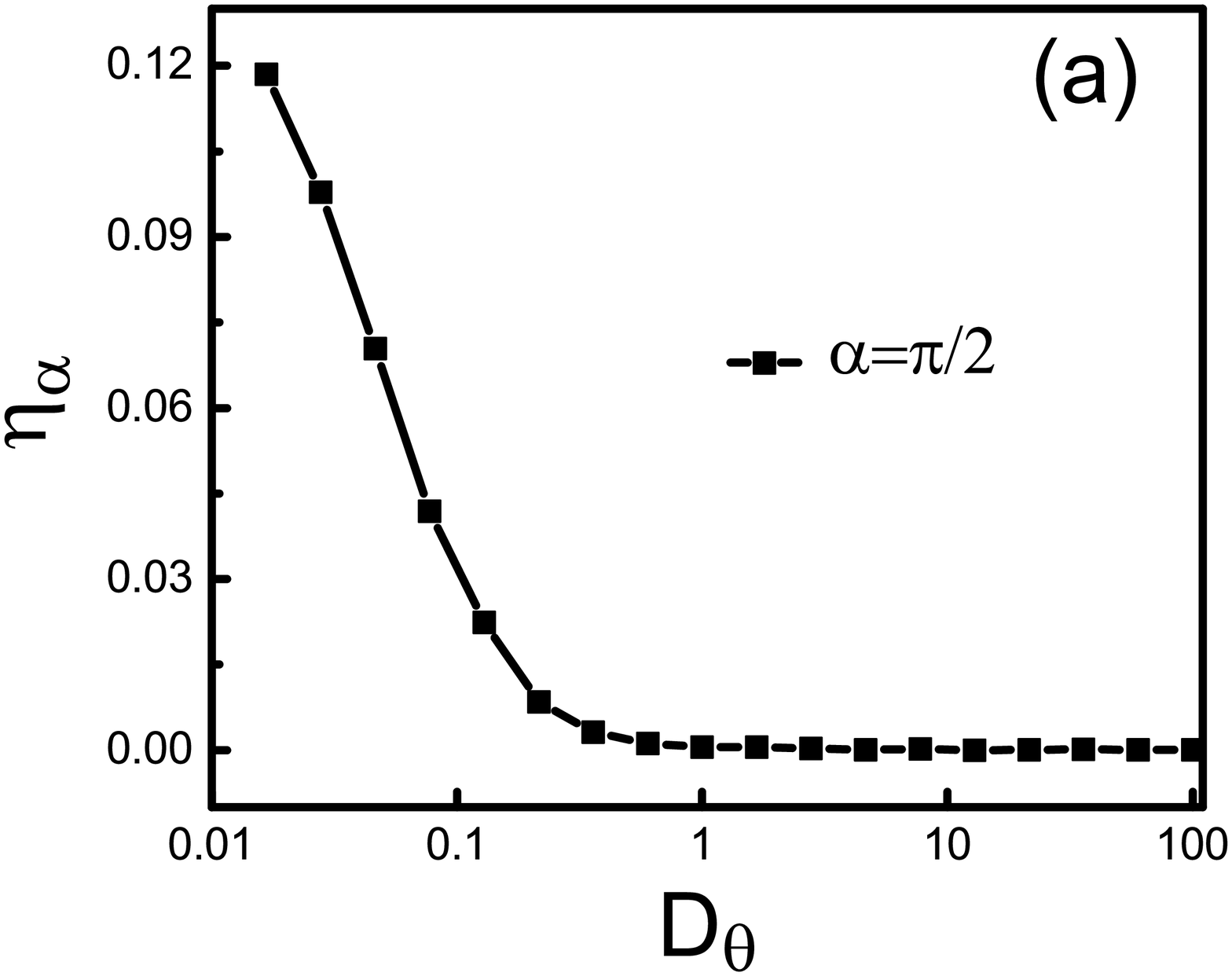}
  \includegraphics[width=0.45\columnwidth]{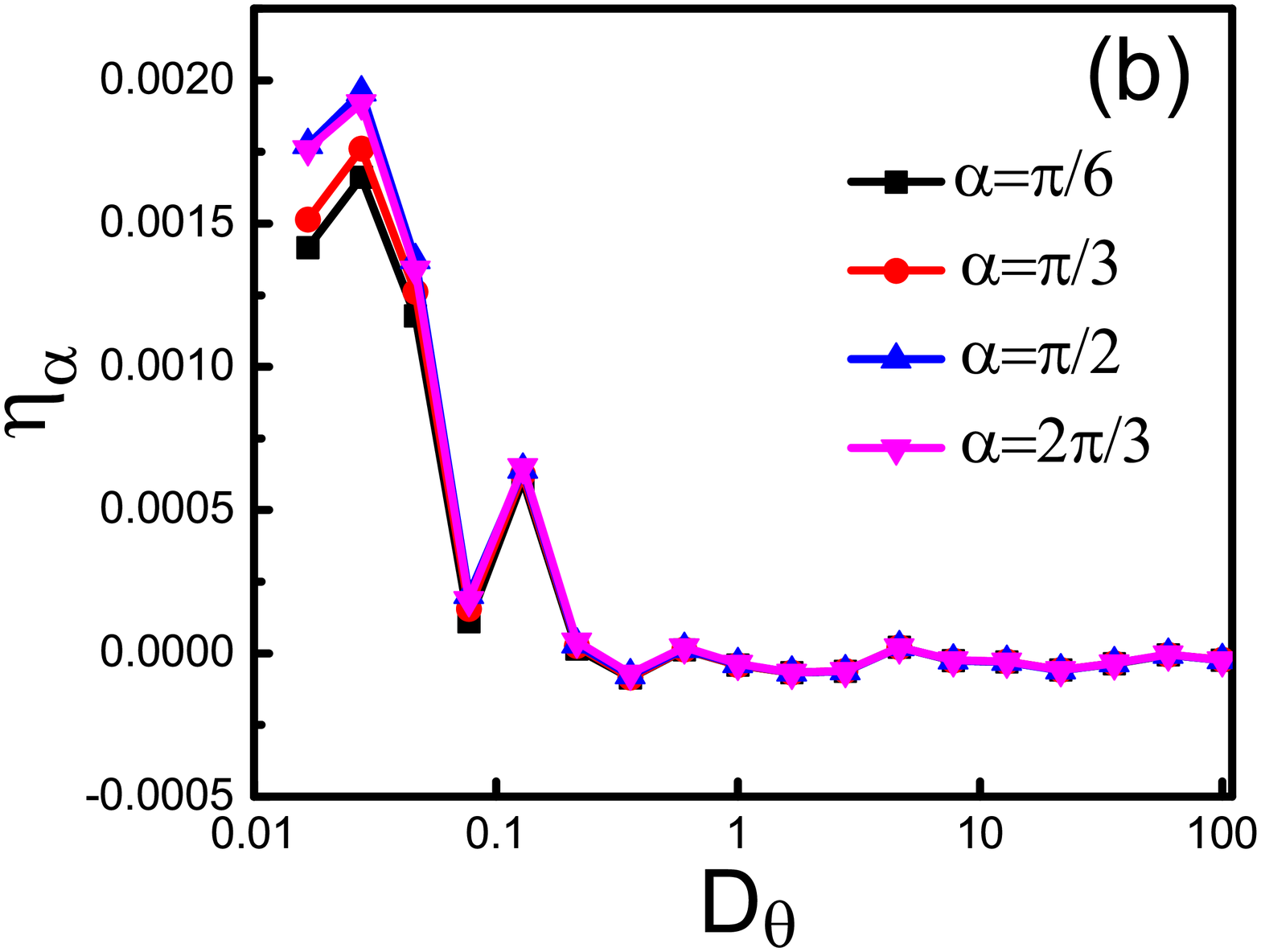}
  \includegraphics[width=0.45\columnwidth]{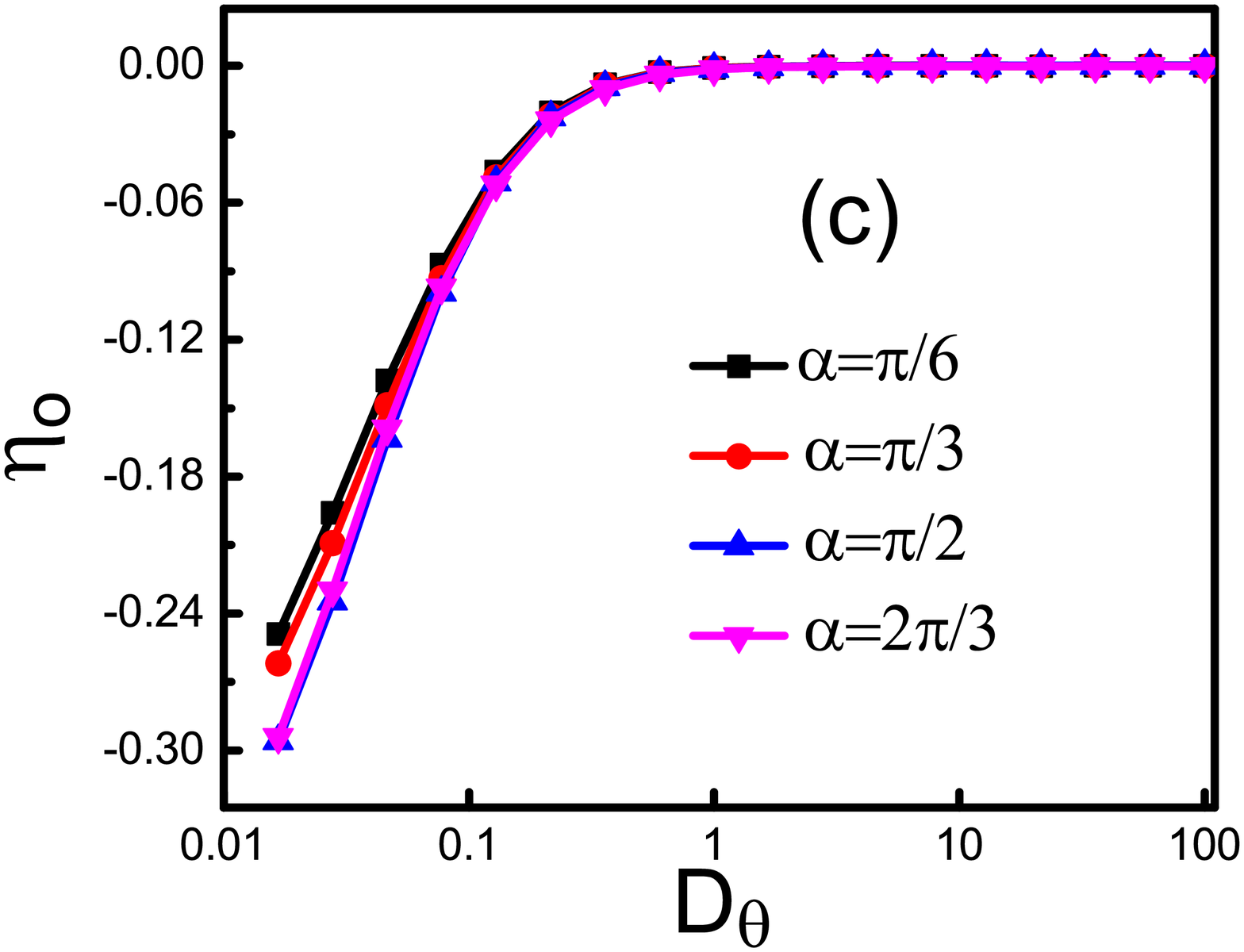}
  \caption{The scaled average velocity $\eta_a$ and $\eta_o$ as a function of the rotational diffusion coefficient $D_\theta$. (a) Active particles for the fixed case at $\alpha=\pi/2$.
  (b) Active particles for the moving case at $\alpha=\pi/6$, $\pi/3$, $\pi/2$, and $2\pi/3$.
  (c) The moving barrier at $\alpha=\pi/6$, $\pi/3$, $\pi/2$, and $2\pi/3$. The other parameters are $\Omega=-0.05$, $D_0=0.1$, $v_0=1.0$, $N=150$, $L_y=16.0$, and $n_p=9$. }
\end{figure}

\indent The dependence of the scaled average velocity on the rotational diffusion coefficient $D_\theta$ is illustrated in Fig. 4. It is found that the curves are similar when the barrier is fixed or can move (shown in Figs. 4(a) and 4(b)).
When $D_\theta \to 0$, the self-propelled angle $\theta$ almost does not
change, and the scaled average velocity approaches its maximal value. As
$D_\theta$ increases to be large, the particles cannot feel the
self-propelled driving and the ratchet effect reduces, so $\eta_a$ and $|\eta_o|$ decreases
and tends to zero. Similarly to the previous figures, the transport speed of
the barrier is much larger than that of active particles
in the both two cases (see Fig. 4(c)). Due to the small scaled average velocity of chiral
particles in the moving case, the curves in Fig.4 (b) are not smooth in the
presence of statistical errors. We can get smoother curves by increasing the number of realizations or the total integration time.
\begin{figure}[htpb]
\vspace{1cm}
  \label{fig5}\includegraphics[width=0.45\columnwidth]{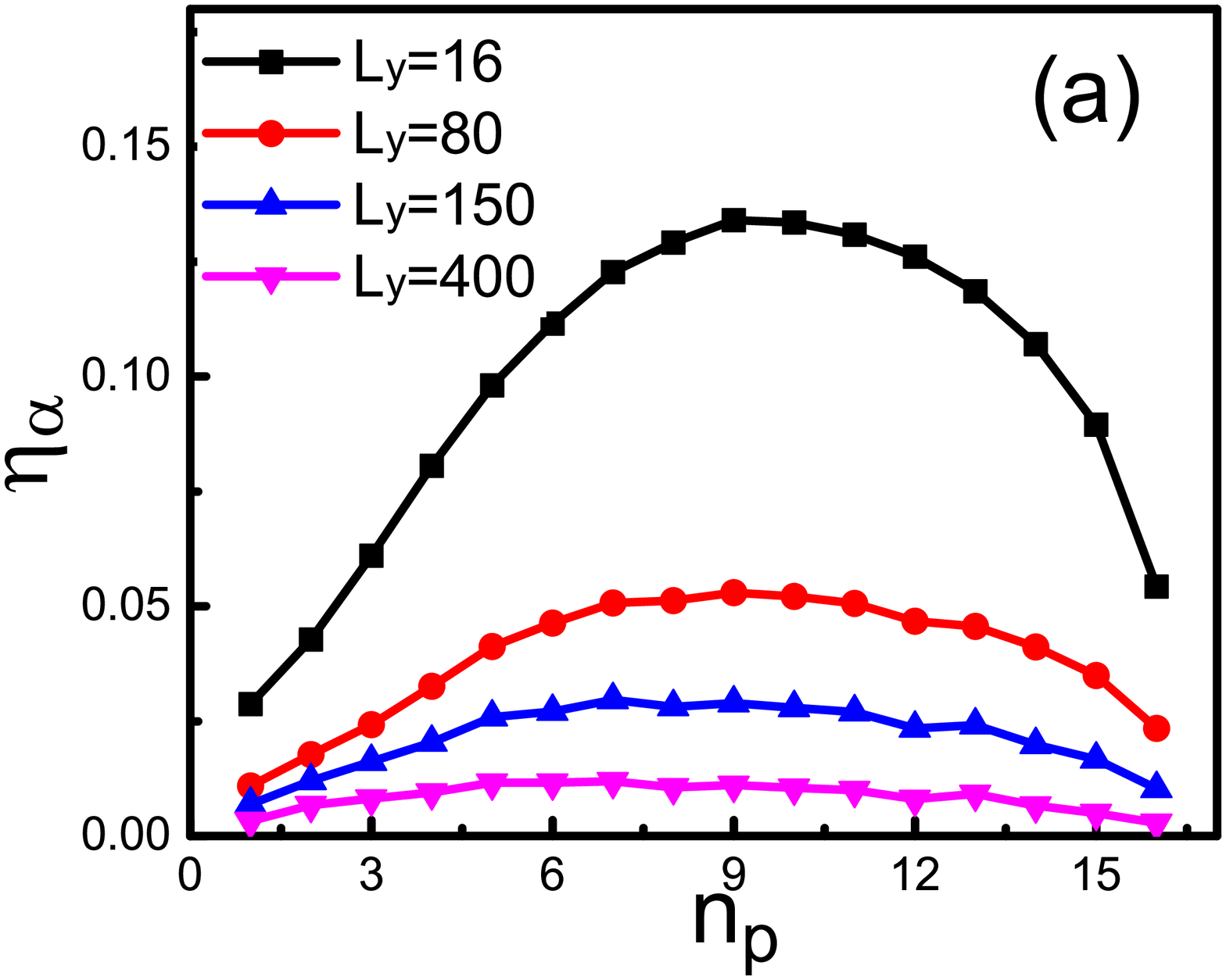}
  \includegraphics[width=0.45\columnwidth]{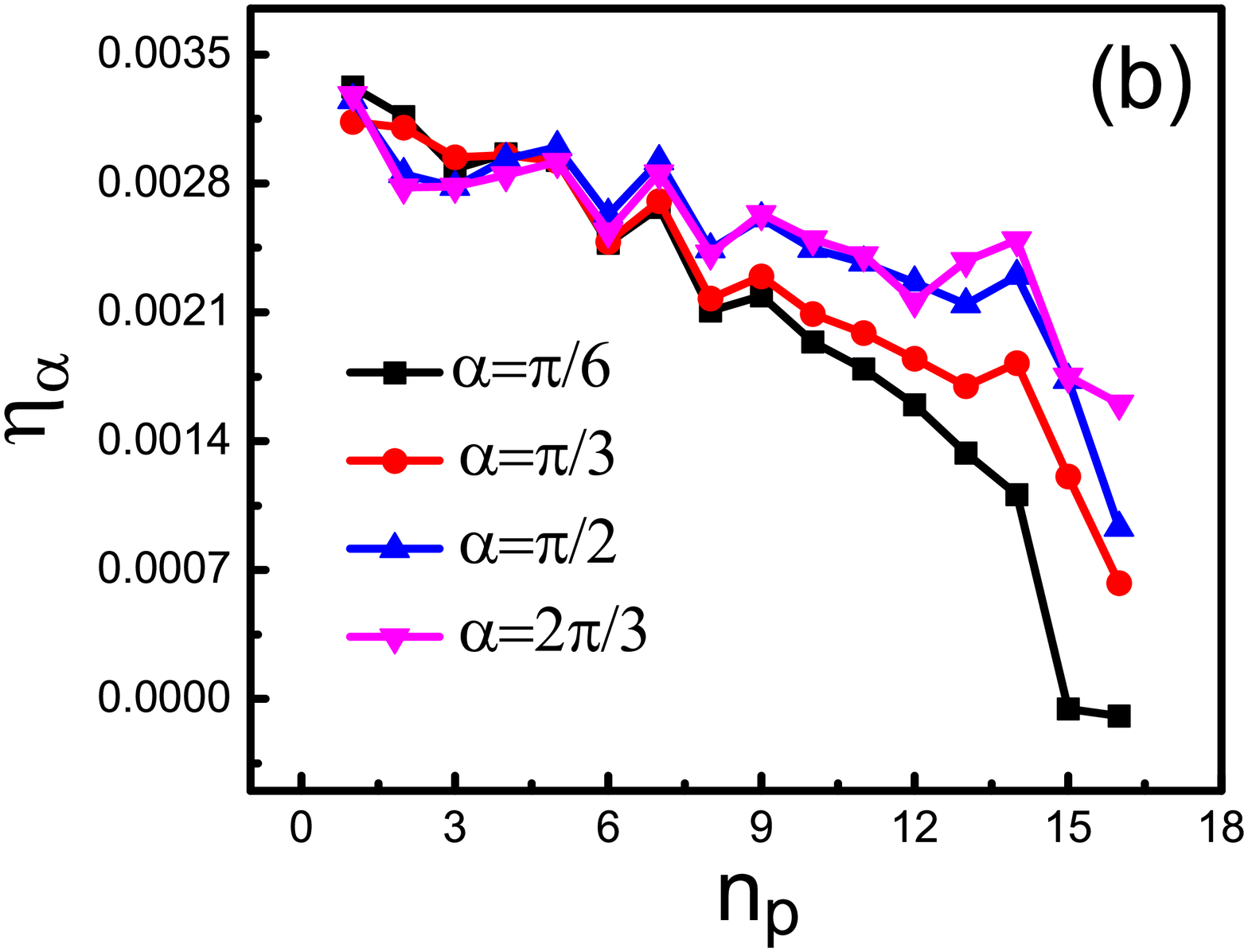}
  \includegraphics[width=0.45\columnwidth]{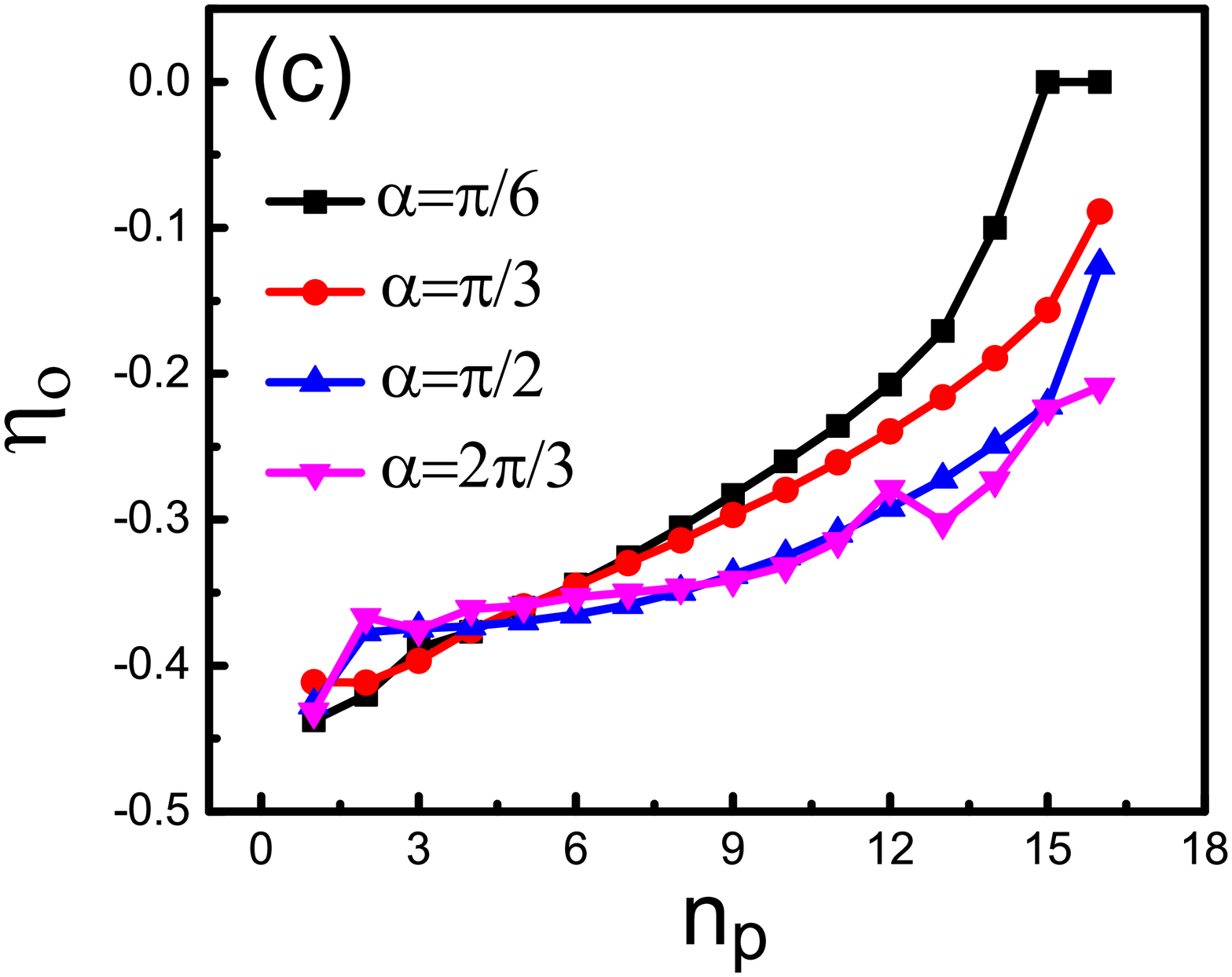}
  \caption{The scaled average velocity $\eta_a$ and $\eta_o$ as a function of the particle number of the V-shaped barrier $n_p$. (a) Active particles for the fixed case at $\alpha=\pi/2$ for different values of the channel widths $L_y$. (b) Active particles for the moving case at $L_y=16$ for different values of the barrier angle $\alpha$. (c) The moving barrier at $L_y=16$ for different values of the barrier angle $\alpha$.
  The other parameters are $\Omega=-0.05$, $D_0=0.1$, $v_0=1.0$, $n_a=140$, and $D_\theta=0.01$. }
\end{figure}

\indent In Figure 5, we present the scaled average velocity as a function of the
particle number $n_p$ of the V-shaped barrier. In the case of the barrier
fixed (see Fig. 5(a)), the curve of active particles is observed to be bell
shaped, and there exists an optimal value of $n_p$ at which $\eta _a$
takes its maximal value. It can be explained as follows. When $n_p$ is very
small, the channel is near to symmetric and the effect of the asymmetric
barrier disappears. The scaled average velocity tends to zero. As $n_p$
increases, the channel becomes asymmetric, and $\eta _a$ increases. When
$n_p$ increases enough to block the channel, the barrier separates the
channel into two parts and particles cannot pass across the barrier, thus
$\eta _a$ goes to zero. Therefore, the optimal $n_p$ can
facilitate the ratchet transport. When increasing the channel widths $L_y$, the magnitude of the scaled average velocity $\eta _a$ decreases because the effect of the asymmetric barrier is reduced. And the position of the optimal number $n_p$ shifts to small $n_p$ slightly. Namely, the position of the optimal number $n_p$ is insensitive to the channel widths $L_y$. When the barrier can move (shown in Figs. 5(b) and 5(c)), for very small $n_p$, a large number of particles act on the barrier which
consists of few particles, resulting in maximal transport speed of the barrier
(see Fig. 5(c)). On increasing $n_p$, because the driving effect decreases,
the scaled average velocity decreases monotonically and finally tends to zero
for large $n_p$.

\begin{figure}[htpb]
\vspace{1cm}
  \label{fig6}\includegraphics[width=0.45\columnwidth]{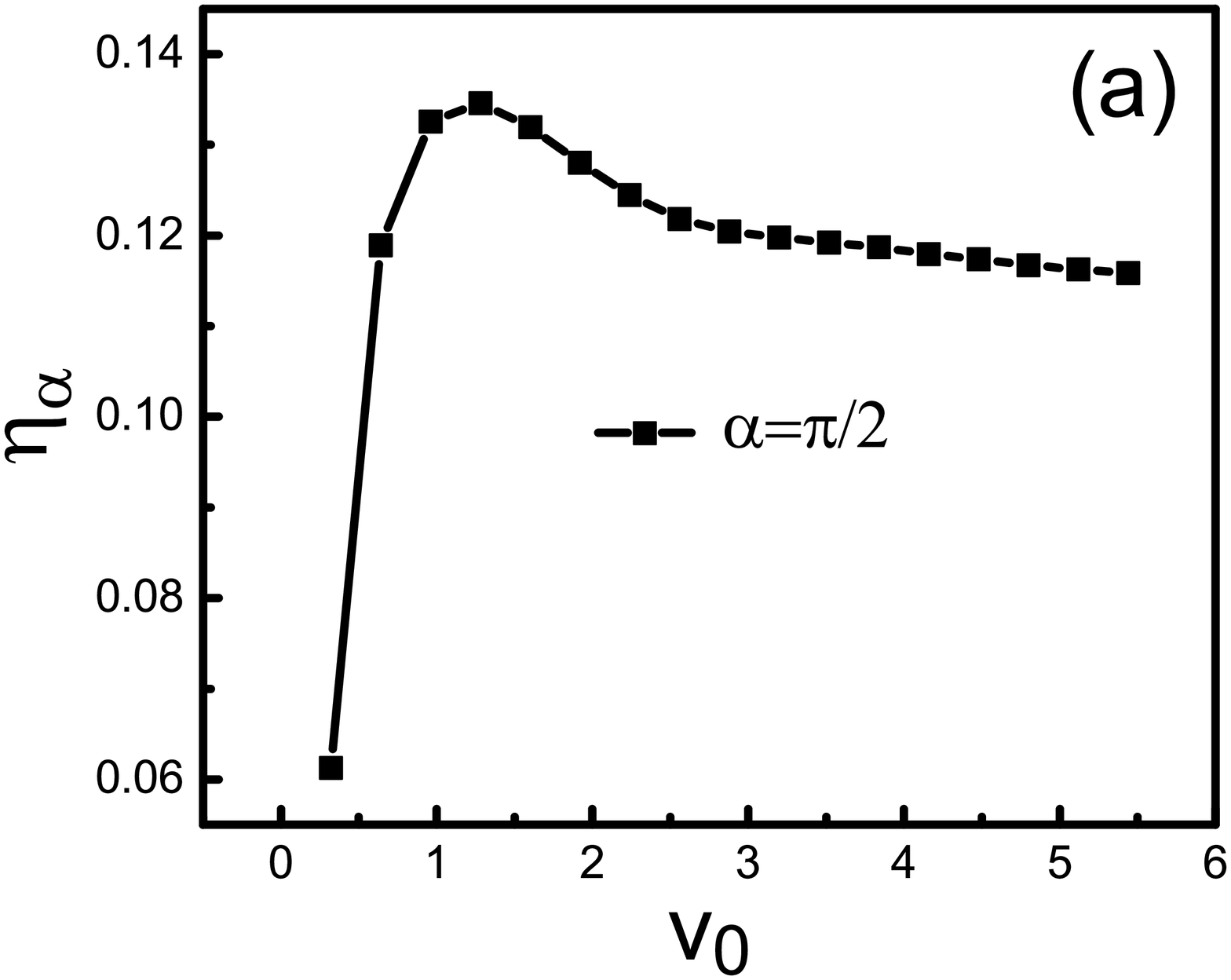}
  \includegraphics[width=0.45\columnwidth]{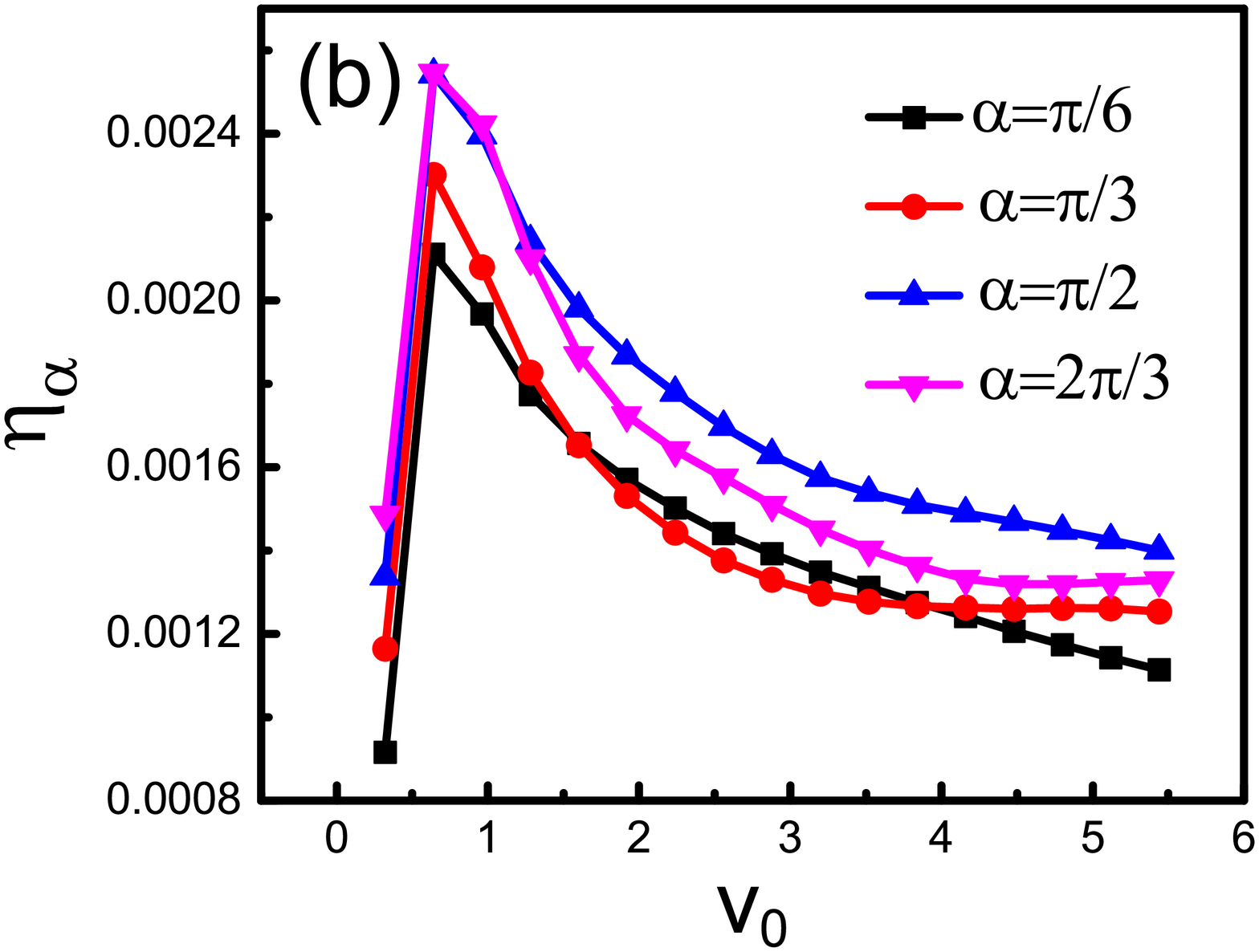}
  \includegraphics[width=0.45\columnwidth]{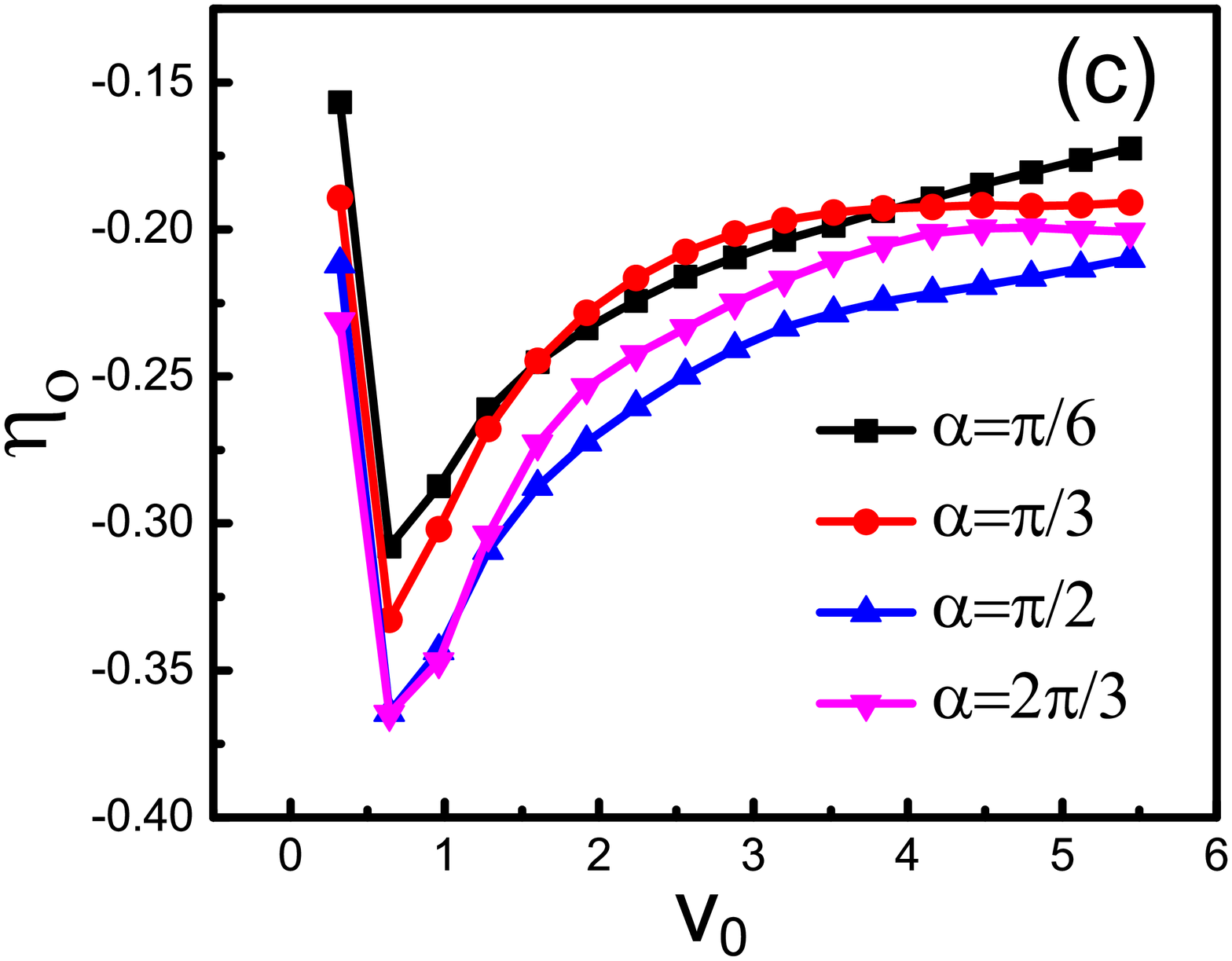}
  \caption{The scaled average velocity $\eta_a$ and $\eta_o$ as a function of the self-propulsion speed $v_0$. (a) Active particles for the fixed case at $\alpha=\pi/2$.
  (b) Active particles for the moving case at $\alpha=\pi/6$, $\pi/3$, $\pi/2$, and $2\pi/3$. (c) The moving barrier at $\alpha=\pi/6$, $\pi/3$, $\pi/2$, and $2\pi/3$.
  The other parameters are $\Omega=-0.05$, $D_0=0.1$, $n_p=9$, $N=150$, $L_y=16.0$, and $D_\theta=0.01$. }
\end{figure}

\indent Figure 6 shows the scaled average velocity versus the self-propulsion speed
$v_0$. For the case of the barrier fixed (see Fig. 6(a)), the fluctuating
input and the ratchet effect disappear as $v_0$ tends to zero,
thus $\eta _a$ is nearly equal to zero. As $v_0$ increases, the
rectification approaches its maximal value. With a further increase in $v_0
$, the scaled average velocity decreases gradually and then tends to a
constant. However when $v_0$ is large enough, the asymmetric effect can be
negligible, thus the scaled average velocity decreases and tends to zero
(not shown in the figure). When the barrier can move (see Fig. 6(b) and
Fig. 6(c)), $\eta _a$ goes to zero as $v_0 \to 0$. For large values of $v_0
$, the asymmetric effect disappears more easily than that in the fixed case
due to the motion of the barrier, thus the directed transport decreases
sharply. Therefore, the optimal self-propulsion speed can facilitate the
rectification of active particles.

\begin{figure}[htpb]
\vspace{1cm}
  \label{fig7}\includegraphics[width=0.45\columnwidth]{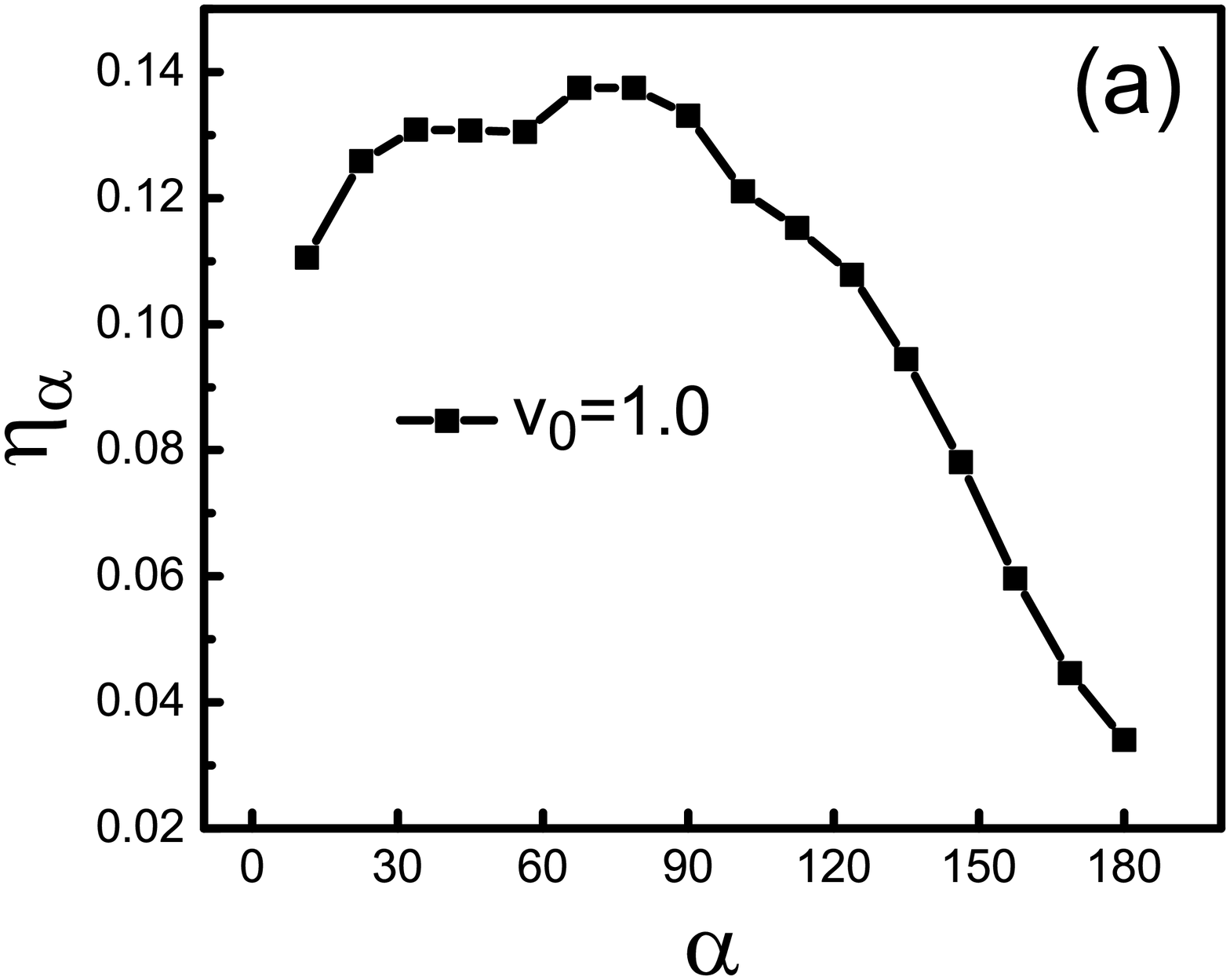}
  \includegraphics[width=0.45\columnwidth]{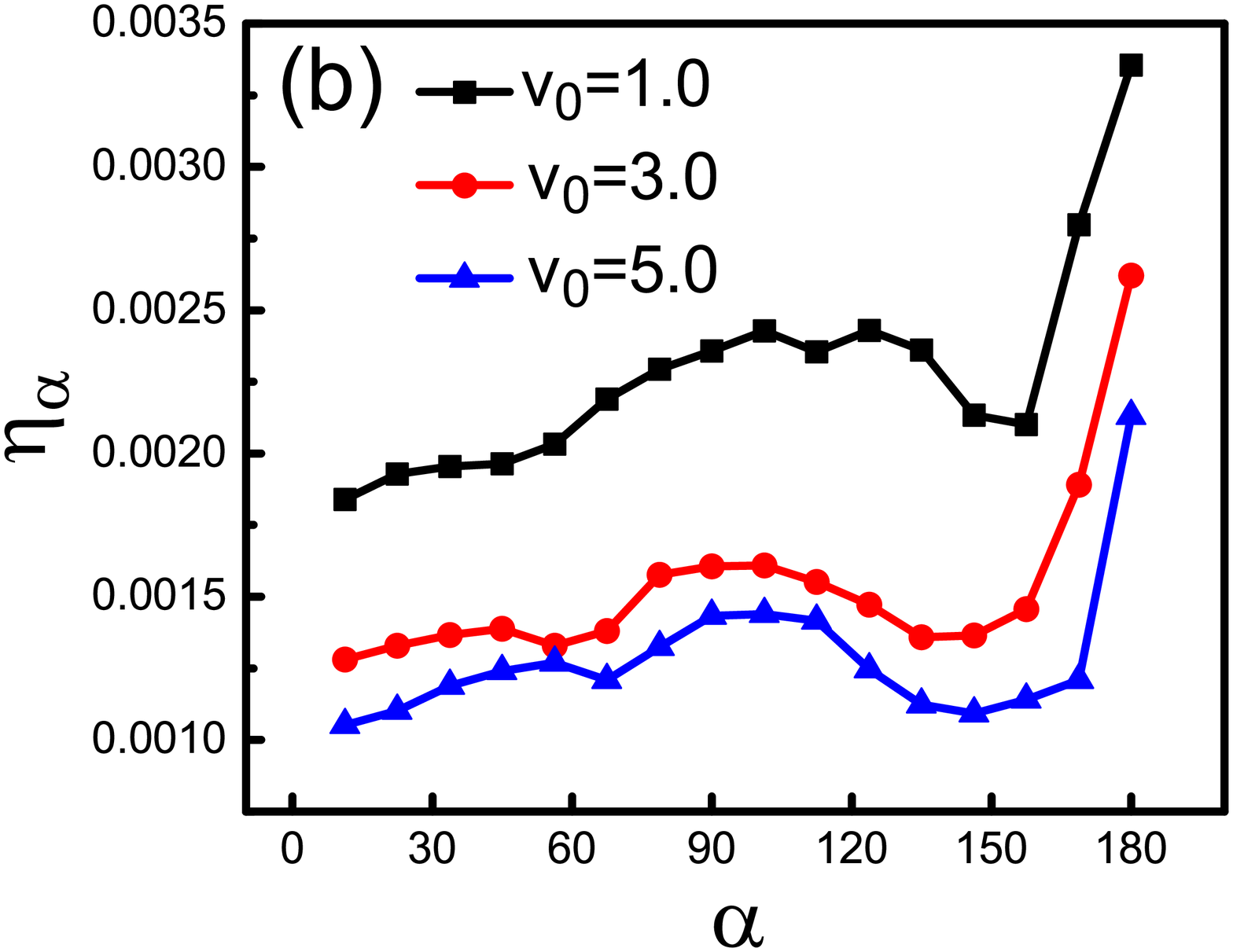}
  \includegraphics[width=0.45\columnwidth]{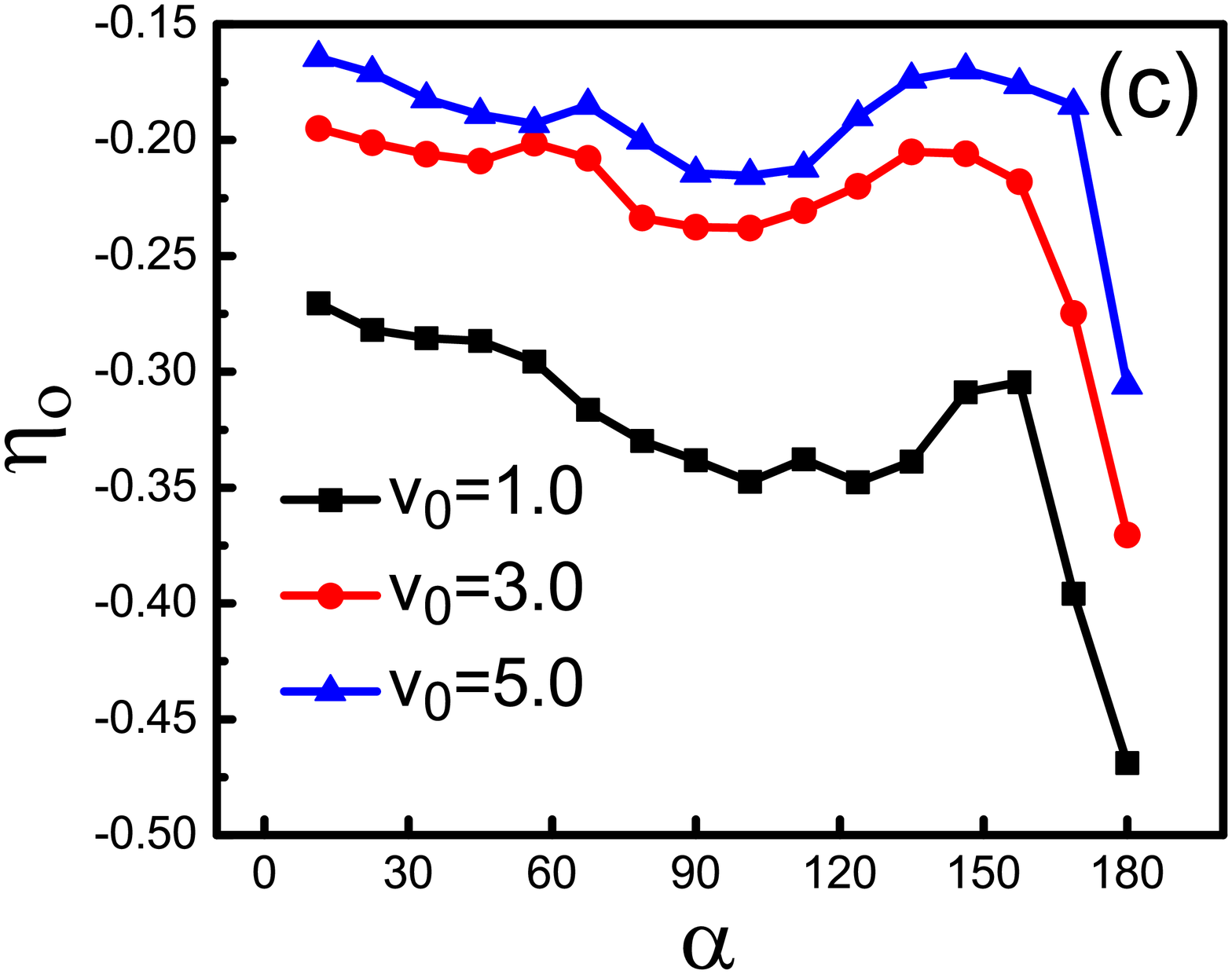}
  \caption{The scaled average velocity $\eta_a$ and $\eta_o$ as a function of the barrier angle $\alpha$. (a) Active particles for the fixed case at $v_0=1.0$.
  (b) Active particles for the moving case at $v_0=1.0$, $3.0$, and $5.0$. (c) The moving barrier at $v_0=1.0$, $3.0$, and $5.0$.
  The other parameters are $\Omega=-0.05$, $D_0=0.1$, $n_p=9$, $ N=150$, $L_y=16.0$, and $D_\theta=0.01$. }
\end{figure}

\indent The dependence of the scaled average velocity on the barrier angle $\alpha$ is shown in Figure 7.
When the barrier is fixed (shown in Fig. 7(a)), there exists an optimal value at which
$\eta _a$ takes its maximal value. When $\alpha \to 0$, particles can pass
through the V-shaped barrier because the height of the V-shaped barrier is
smaller than the height of channel $L_y$ and the channel is not
blocked. Thus, $\eta _a$ is small but does not tend to zero. For very large
value of $\alpha $, the asymmetry effect disappear and no directed
transport occurs, thus $\eta _a\rightarrow0$. When the barrier can move (shown in Figs. 7(b) and 7(c)), $\eta _a$ and $|\eta_o|$ increase slightly with
an increase in $\alpha $. In other words, the scaled average velocity is
insensitive to the angle $\alpha$. This is consistent with the other figures in the moving case. In particular, the scaled average velocity
reaches to the maximum in the limit case $\alpha =\pi$. We can explain as
follows: when $\alpha \ne \pi $, the V-shaped barrier has two sides and each
side has $n_p$ particles. The driving forces act on every particle of each
side from both the positive and negative direction of $x$. When $\alpha =\pi
$, the pushing effect on the barrier which becomes a straight stick is
bigger than that in the case of $\alpha \ne \pi $. Because the average force exerted by active particles is mainly along the negative direction of $x$. Thus, maximal rectification is achieved when $\alpha =\pi$.

\begin{figure}[htpb]
\vspace{1cm}
  \label{fig8}\includegraphics[width=0.45\columnwidth]{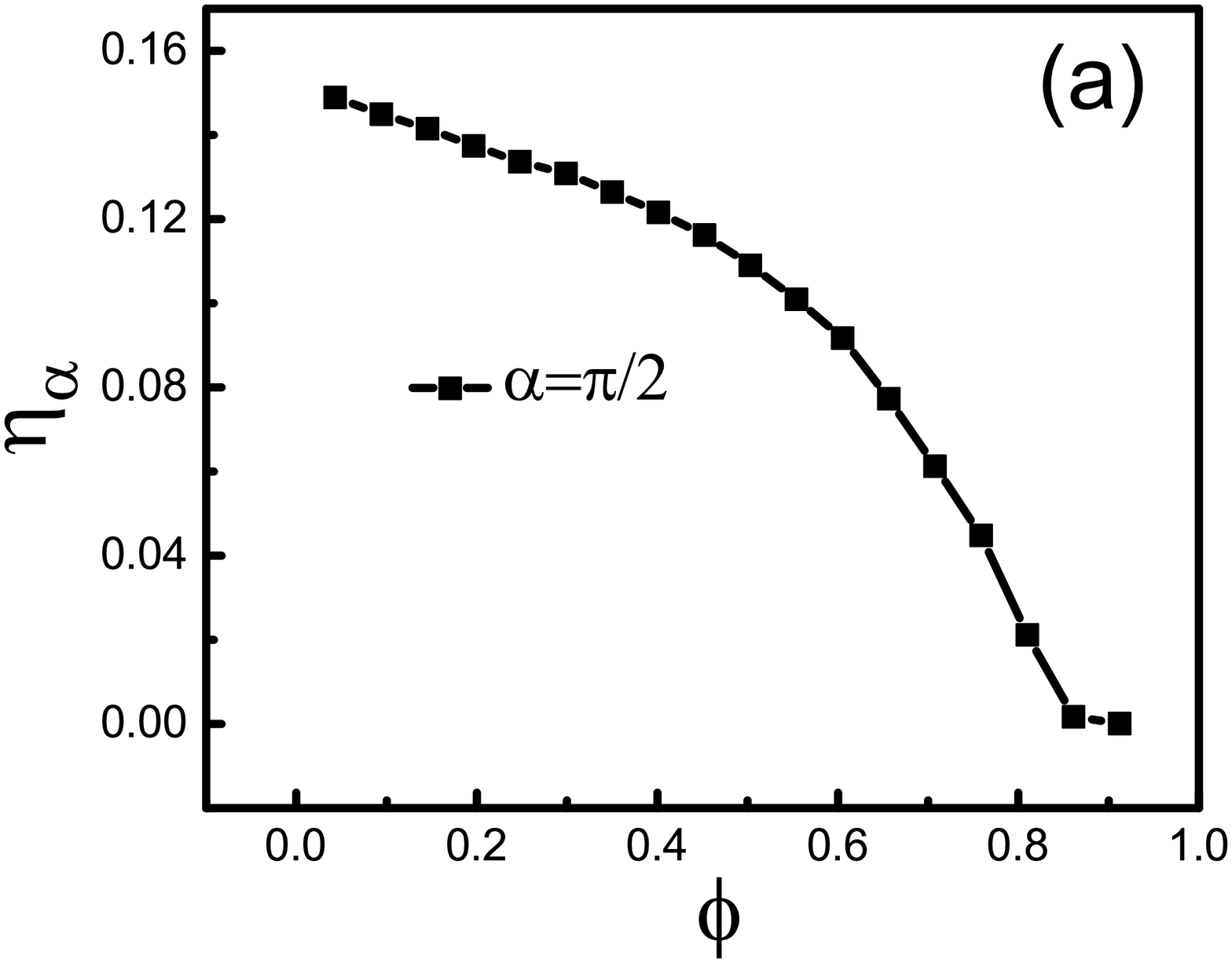}
  \includegraphics[width=0.45\columnwidth]{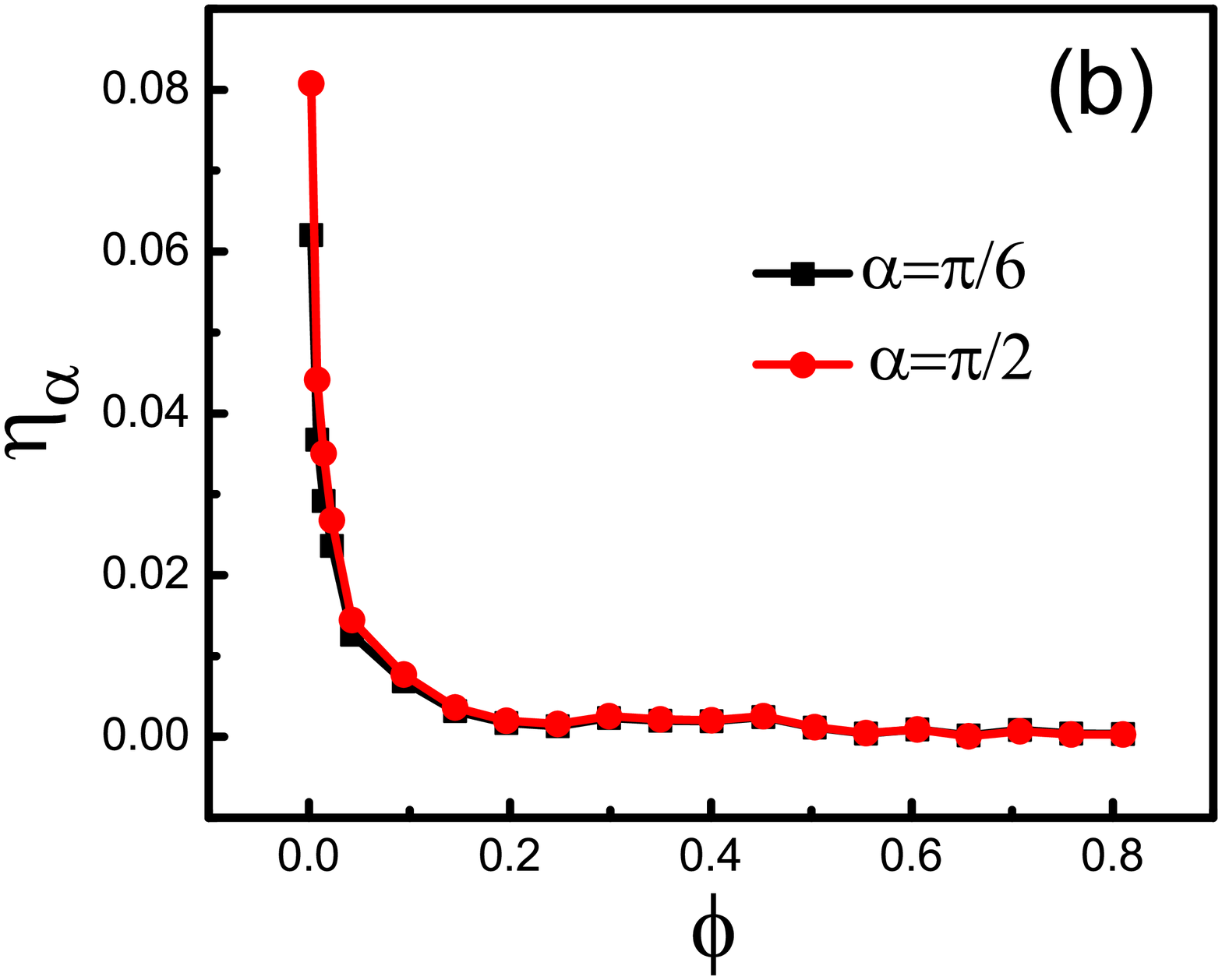}
  \includegraphics[width=0.45\columnwidth]{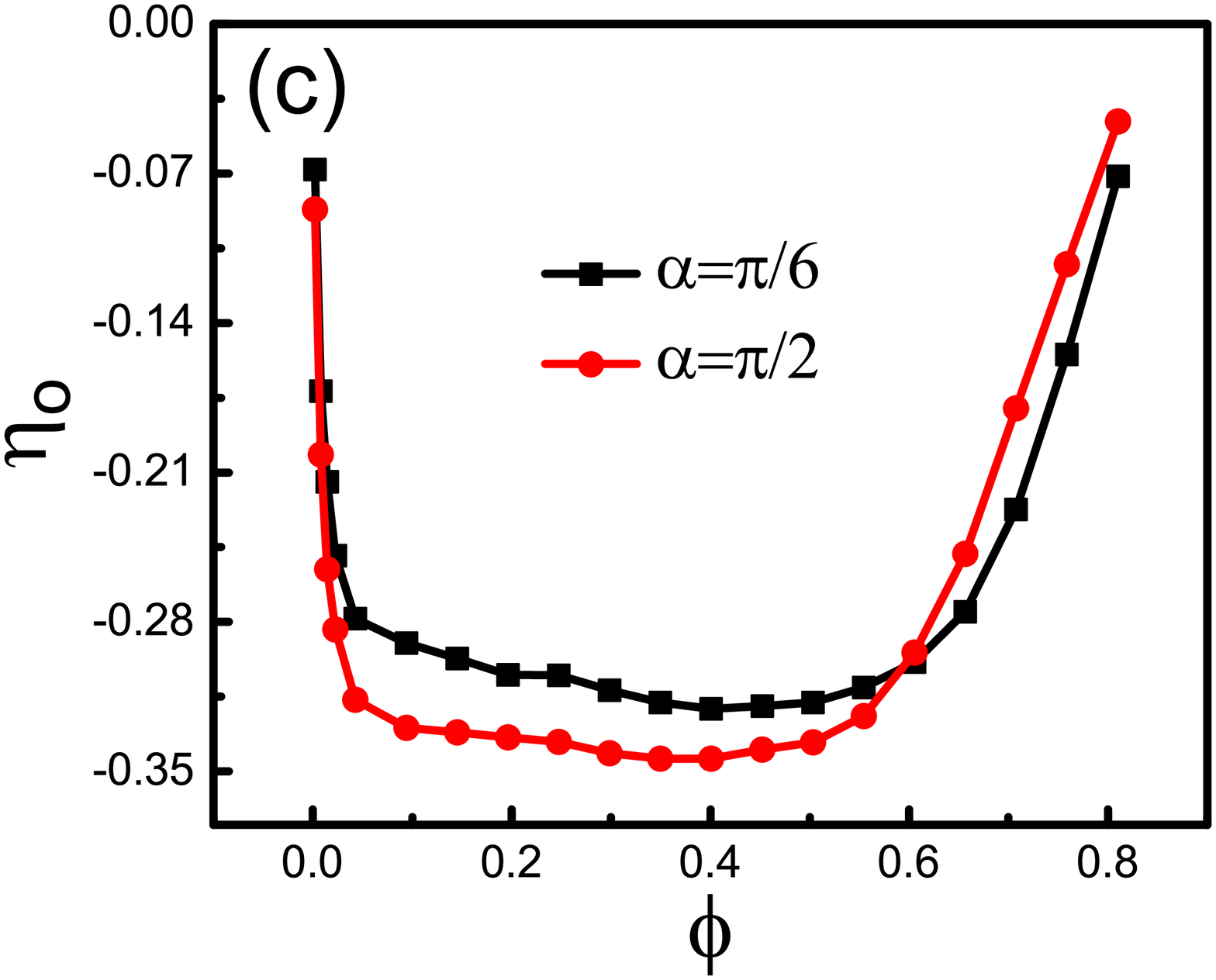}
  \caption{The scaled average velocity $\eta_a$ and $\eta_o$ as a function of the packing fraction $\phi$. (a) Active particles for the fixed case at $\alpha=\pi/2$.
  (b) Active particles for the moving case at $\alpha=\pi/6$ and $\pi/2$. (c) The moving barrier at $\alpha=\pi/6$ and $\pi/2$.
  The other parameters are $\Omega=-0.05$, $D_0=0.1$, $v_0=1.0$, $n_p=9$, $L_y=16.0$, and $D_\theta=0.01$. }
\end{figure}

\begin{figure}[htpb]
\vspace{1cm}
  \label{fig9}\includegraphics[width=0.45\columnwidth]{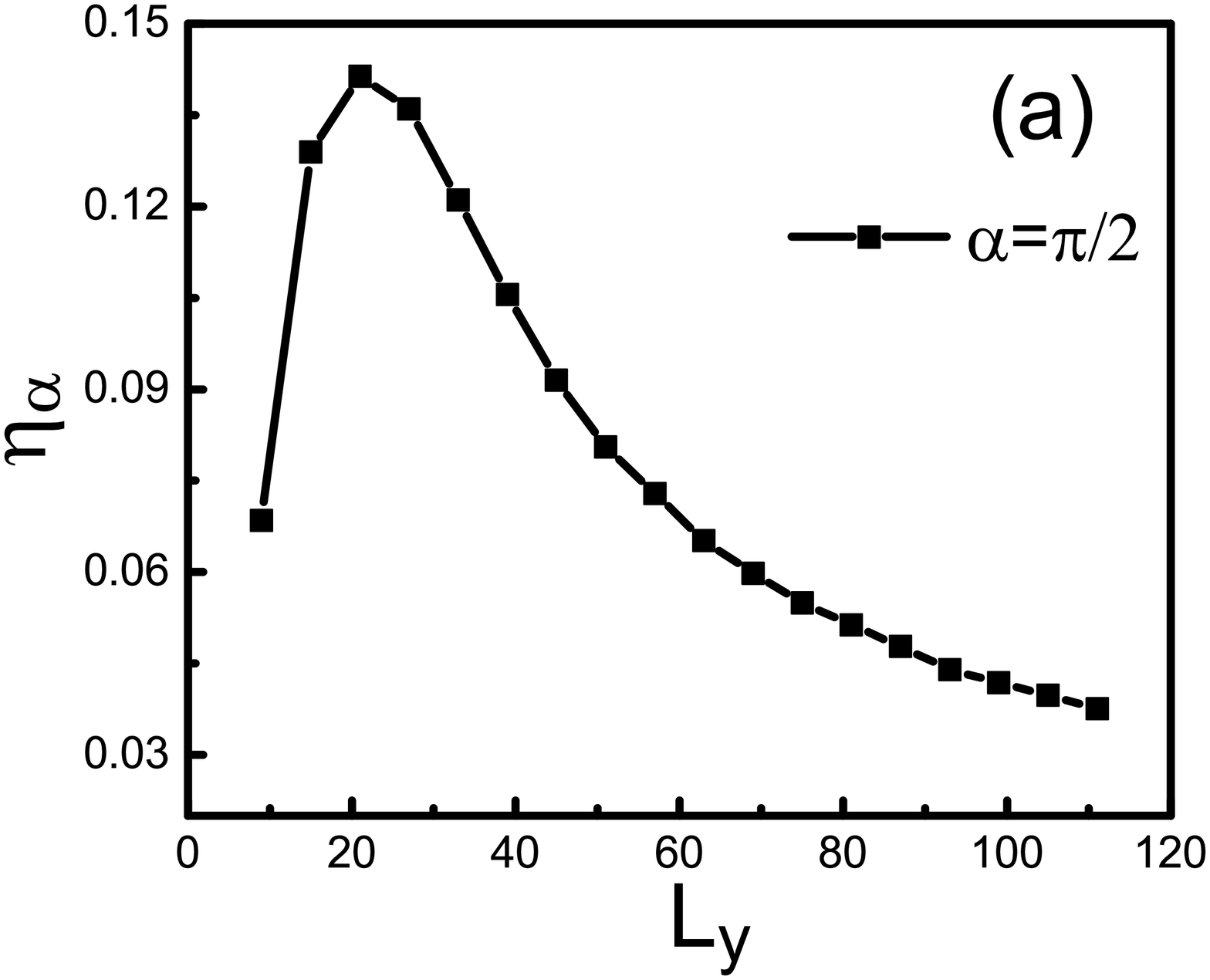}
  \includegraphics[width=0.45\columnwidth]{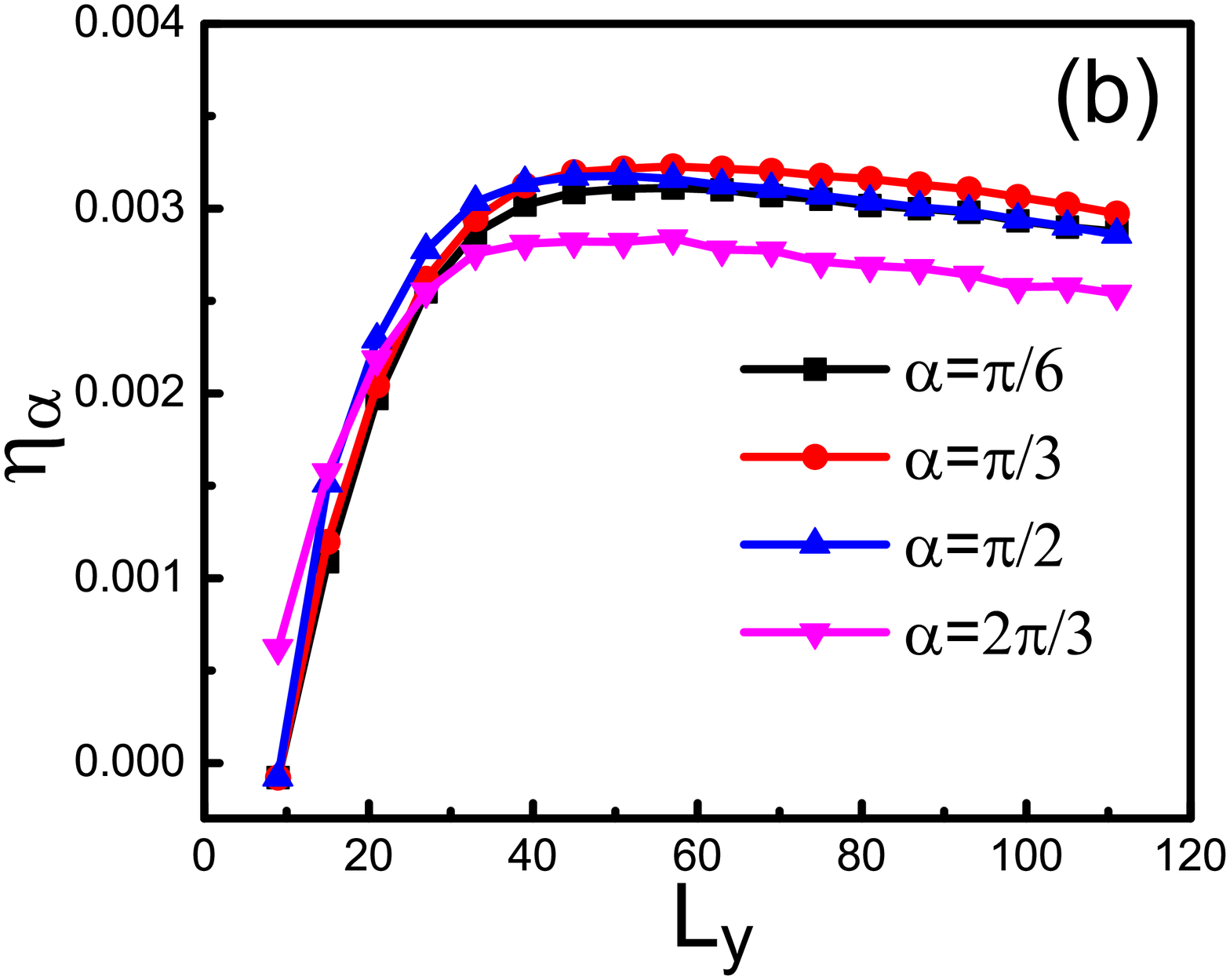}
  \includegraphics[width=0.45\columnwidth]{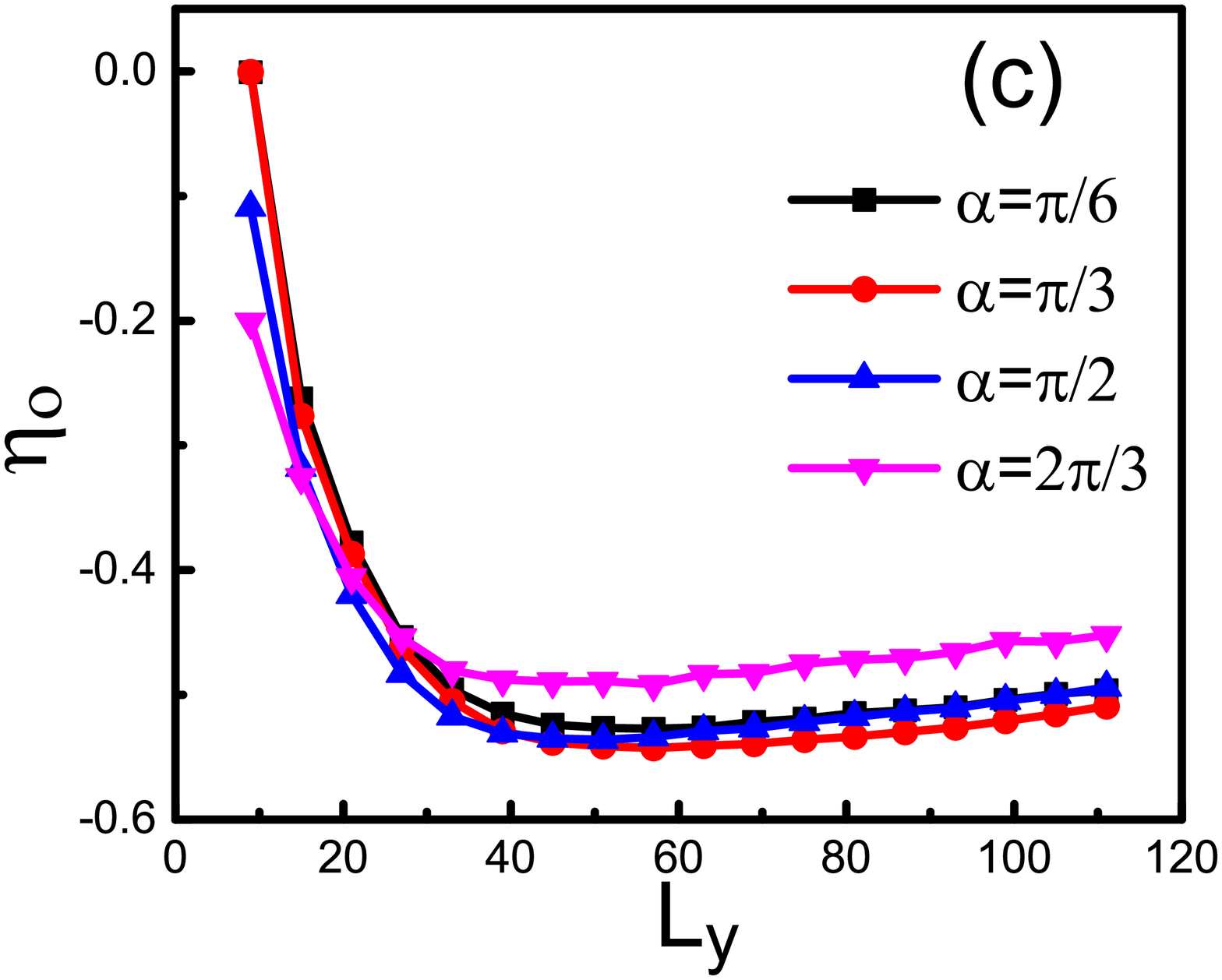}
  \caption{The scaled average velocity $\eta_a$ and $\eta_o$ as a function of the channel width $L_y$. (a) Active particles for the fixed case at $\alpha=\pi/2$.
  (b) Active particles for the moving case at $\alpha=\pi/6$, $\pi/3$, $\pi/2$, and $2\pi/3$. (c) The moving barrier at $\alpha=\pi/6$, $\pi/3$, $\pi/2$, and $2\pi/3$.
  The other parameters are $\Omega=-0.05$, $D_0=0.1$, $v_0=1.0$, $n_p=9$, $n=150$, and $D_\theta=0.01$. }
\end{figure}

\indent Figure 8 depicts the scaled average velocity as a function of the packing fraction
$\phi$. When the barrier is fixed (see Fig. 8(a)), the rectification of active particles
decreases slowly with the increasing $\phi$. For a large $\phi $, the
particles are jammed, thus $\eta _a$ tends to zero. When the barrier can move (shown in Figs. 8(b) and 8(c)), the rectification of active particles decreases sharply with
the increasing $\phi$ because the driving effect increases sharply (see Fig. 8(b)).
$\eta _a $ also tends to zero as $\phi \to 1$ due to the jammed particles. For the moving barrier (see Fig. 8(c)),
the pushing effect is small since only few active particles are contributing
when $\phi $ tends to zero, thus $|\eta_o| \to 0$. For high concentration, the active bath is
jammed, which leaves no mobility for the barrier, so $|\eta_o| \to 0$. Therefore, there exists an
optimal packing fraction that maximizes the scaled average velocity of the barrier. As the above results, the velocity ratio between the barrier and active particles is decided by the number of active particles.

\indent The dependence of the scaled average velocity on the channel width $L_y$ is shown in Fig. 9. In the present system, the arm length of the barrier is 9 and $L_y$ must be larger than 9. In the case of the barrier fixed (see Fig. 9(a)), there exists an optimal value of $L_y$ at which $\eta _a$ takes its maximal value. It can be explained as follows. When $L_y$ is very small, the barrier blocks the channel, particles cannot pass across the barrier, thus $\eta _a$ goes to zero. When $L_y \rightarrow \infty$, the channel is near to symmetric and the effect of the asymmetric barrier disappears, thus, $\eta_a \rightarrow 0$. When the barrier can move (shown in Figs. 9(b) and 9(c)), for very small $L_y$, particles are difficult to stride over the obstacle, thus $\eta _a$ and $|\eta_o|$ tends to zero. On increasing $L_y$, $\eta _a$ and $|\eta_o|$ increase monotonically and reach the maximum because particles cross the barrier more and more easily. However, when $L_y \rightarrow \infty$, most of particles do not interact with the barrier and particle-barrier interaction becomes insignificant. Therefore, the ratchet effect disappears, $\eta _a$ and $|\eta_o|$ tends to zero (not shown in the figure).

\indent Finally, we discuss the possibility of realizing our model in experimental setups. Consider a system of Bacillus subtilis (diameter about 1 $\mu m$) moving in a two-dimensional channel at room temperature. The suspension of bacteria is grown for 8-12h in Terrific Broth growth medium (Sigma Aldrich). We can continuously measure the optical scattering of the medium using an infrared proximity sensor to monitor the concentration of bacteria during the growth phase \cite{Kaiser}. A V-shaped barrier is fabricated by photolithography \cite{ref55,ref56}. To control the orientation of the barrier with an external magnetic field, we can mix a liquid photoresist SU-8 with micron-size magnetic particles before spin coating \cite{Kaiser}. In order to restrict the V-shaped barrier moving only along the $x$-direction, two parallel tracks (active particles cannot feel the tracks) are settled in the channel, one is fixed at the bottom of channel and the other is fixed at the top of the barrier. The influence of gravity is negligible. Due to the chirality of active particles and the transversal asymmetry of the barrier position, active particles can power and steer the directed transport of the barrier in the longitudinal direction. The motion of the barrier and active particles are captured by a digital high-resolution microscope camera, from which the average velocity can be calculated.

\section{Concluding Remarks}
\indent In conclusion, we numerically studied the transport of a moving V-shaped barrier exposed to a bath of chiral
active particles in a two-dimensional channel. It is found that the barrier can be driven to move directly along the bottom
of the channel by chiral active particles. When the barrier is fixed at the
bottom of the channel, the upper-lower asymmetric due to the position of the V-shaped
barrier and the intrinsic property of chiral particles can break
thermodynamical equilibrium and induce the rectified transport of active
particles. Chiralities determine the transport direction of active
particles. By choosing suitable system parameters, the transport efficiency of active
particles can reach the maximum. When the V-shaped barrier can move along the bottom
of the channel, the nonequilibrium driving which comes from the chiral
particles breaks thermodynamical equilibrium and power the barrier to move
in the $x$-direction. The transport of the barrier is determined by the chirality of active particles. The moving barrier and active particles move in the opposite directions. Comparing the transport of chiral active particles between the cases of the barrier fixed
and moving, the rectified efficiency of active particles in the
moving case is reduced much than that in the fixed case. The velocity ratio between the barrier and active particles is decided by the number of active particles. Maximal transport
velocities of active particles and the barrier are obtained when the system
parameters are optimized. In particular, changing $n_p$, $\alpha$, and
$\phi$ in the moving case lead to the transport behaviors more different
than that in the fixed case. In other words, tailoring the geometry of the
barrier and the active concentration provides novel strategies to control the transport properties of
micro-objects or cargoes in an active medium. Maybe our results can be
applied practically in propelling carriers and motors by a bath of bacteria
or artificial microswimmers, such as hybrid micro-device engineering, drug delivery, micro-fluidics and lab-on-chip technology.

\section*{Acknowledgments}
This work was supported in part by the National Natural Science Foundation of
China (Grant Nos. 11575064, 61762046 and 11205044), the Natural Science Foundation of Guangdong Province, China (Grant
No. 2014A030313426 and No. 2017A030313029), the Natural Science Foundation of Jiangxi Province,
 China (Grant No. GJJ161580 and No. GJJ160624) and the Innovation Project of Graduate School of South China Normal University.

\end{document}